\documentclass[12pt]{article}

\usepackage[a4paper,left=30mm,right=20mm,top=20mm,bottom=20mm]{geometry}
\usepackage{graphics}
\usepackage{amsmath}
\usepackage{epsfig}
\usepackage{cite}
\usepackage{xcolor}
\usepackage{colordvi}
\usepackage{color}
\usepackage[english]{babel}

\graphicspath{{figures/}}

\title{The role of initial gluon emission in double $J/\psi$ production at central rapidities}

\author{S.P.~Baranov$^1$, A.V.~Lipatov$^{2,3}$, A.A.~Prokhorov$^{2,3,4}$ }

\begin{document}

\maketitle

\begin{center}

{\it $^{1}$P.N.~Lebedev Institute of Physics, Moscow 119991, Russia}\\
{\it $^{2}$Skobeltsyn Institute of Nuclear Physics, Lomonosov Moscow State University, 119991, Moscow, Russia}\\
{\it $^{3}$Joint Institute for Nuclear Research, 141980, Dubna, Moscow region, Russia}\\
{\it $^{4}$Faculty of Physics, Lomonosov Moscow State University, 119991 Moscow, Russia}\\

\end{center}

\vspace{0.5cm}

\begin{center}

{\bf Abstract }

\end{center}

\indent
We consider the process of double $J/\psi$ production in $pp$ collisions at the LHC
in the framework of $k_{T}$-factorization approach. We focus on the gluon fragmentation
mechanism which is related to multiple gluon emission in the initial state. The initial
state emission is treated according to Catani-Ciafaloni-Fiorani-Marchesini evolution 
equation, and applies to both single and double parton scattering cases. 
We show the importance of fragmentation contributions to $J/\psi$ pair production 
and perform a comparison between theoretical predictions and the latest ATLAS data 
collected at $\sqrt{s} = 8$ TeV.
We find that the effects of multiple gluon emission are essential for both single- 
and multiparton interaction processes.
Finally, we highlight the problem of correct choice of the factorization scale
with respect to numerical stability of the calculations and the consistency with
non-collinear evolution equations.

\vspace{1.0cm}


\newpage
\section{Introduction} \indent

The production of $J/\psi$ pairs at high energies serves as an important probe testing the
quarkonia production mechanisms and their interpretation within nonrelativistic quantum
chromodynamics (NRQCD) \cite{1,2,3}. NRQCD provides a rigorous theoretical framework 
commonly used to describe the production and decay of heavy quark bound states.
It implies a factorizable separation between perturbatively calculated short distance 
cross sections for the production of a heavy quark pair and its subsequent 
nonperturbative transition into a physical particle.
The intermediate $Q\bar{Q}$ state $|^{2S+1}L^{(a)}_J\rangle$ is characterized with its 
spin $S$, orbital angular momentum $L$, total angular momentum $J$, and color 
representation $a$. Its transition to a physical meson proceeds via soft gluon
radiation and is described (parametrized) by the long-distance non-perturbative matrix 
elements (LDMEs), which obey certain hierarchy in powers of the relative heavy quark 
velocity $v$ \cite{1,2,3}. Combined with next-to-leading order (NLO) short-distance 
cross sections, NRQCD fits the LHC data on the prompt $J/\psi,\, \psi$ and $\chi_c$ 
transverse momentum distributions (see, for example, \cite{4,5,6,7,8,9,10,11}).
A long-standing challenge in explaining the polarization phenomena (the so-called 
"polarization puzzle") have been solved recently \cite{12} (see also discussions 
\cite{13,14,15}).

In the last few years, NRQCD has made significant progress in evaluating the prompt 
double $J/\psi$ production. A complete leading-order (LO) calculation including 
color singlet (CS) and color octet (CO) terms is done \cite{16}. 
Relativistic corrections to the $J/\psi$ pair production are carried out \cite{17}.
Full NLO contribution to the CS mechanism is known \cite{18}, and partial tree-level 
NLO* contributions to the CS and CO mechanisms are calculated \cite{19}. The latter are 
found to be essential for both low and large transverse momenta, as compared to the LO 
results. 
However, these predictions still suffer from sizeable discrepancies with the latest ATLAS 
data \cite{20}, especially at large transverse momentum $p_T(J/\psi,J/\psi)$, large 
invariant mass $m(J/\psi,J/\psi)$, and large rapidity separation $\Delta y(J/\psi,J/\psi)$
in the $J/\psi$ pairs, that motivates looking for additional mechanisms contributing 
to the double $J/\psi$ events.

In our previous publication \cite{21},
we revealed a sizeable combinatorial contribution from multiple gluon radiation 
originating from the QCD evolution of the initial gluon cascade. This contribution
can be efficiently taken into account using the Ciafaloni-Catani-Fiorani-Marchesini 
(CCFM) evolution equation \cite{22,23,24,25}.
Indeed, gluons emitted in a non-collinear evolution cascade have non-zero transverse 
momenta and give rise to physical $J/\psi$ mesons via color octet fragmentation.
The impact of such processes on double $J/\psi$ production at forward rapidities has 
been investigated \cite{21}, and their importance for the associated $Z/W^{\pm} + J/\psi$ 
production at the LHC has been pointed out \cite{26}.

It has been shown \cite{21}
that the gluon and quark fragmentation into charmonium states could especially play 
a crucial role in the kinematical region of large invariant masses $m(J/\psi,J/\psi)$ 
and/or large rapidity separation $\Delta y(J/\psi,J/\psi)$ between the $J/\psi$ mesons. Given the fact that this
region is covered by the ATLAS experiment \cite{20}, we formulate our next goal.
In our present study, we are going to investigate the effects of multiple gluon 
radiation with respect to double $J/\psi$ production at central rapidities and to give 
a quantitative comparison of our predictions with the available ATLAS data \cite{20}.
The effect of multiple gluon radiation in DPS events is to be studied for the first time.

The outline of the paper is the following. In Section~2 we briefly describe the basic 
steps of our calculations. 
Section~3 is devoted to a discussion on the different choices of the factorization scale.
Section~4 displays our numerical results. Our conclusions are summarised in Section~5.

\section{The model} \indent

\begin{figure}
\begin{center}
{\includegraphics[width=0.8\textwidth]{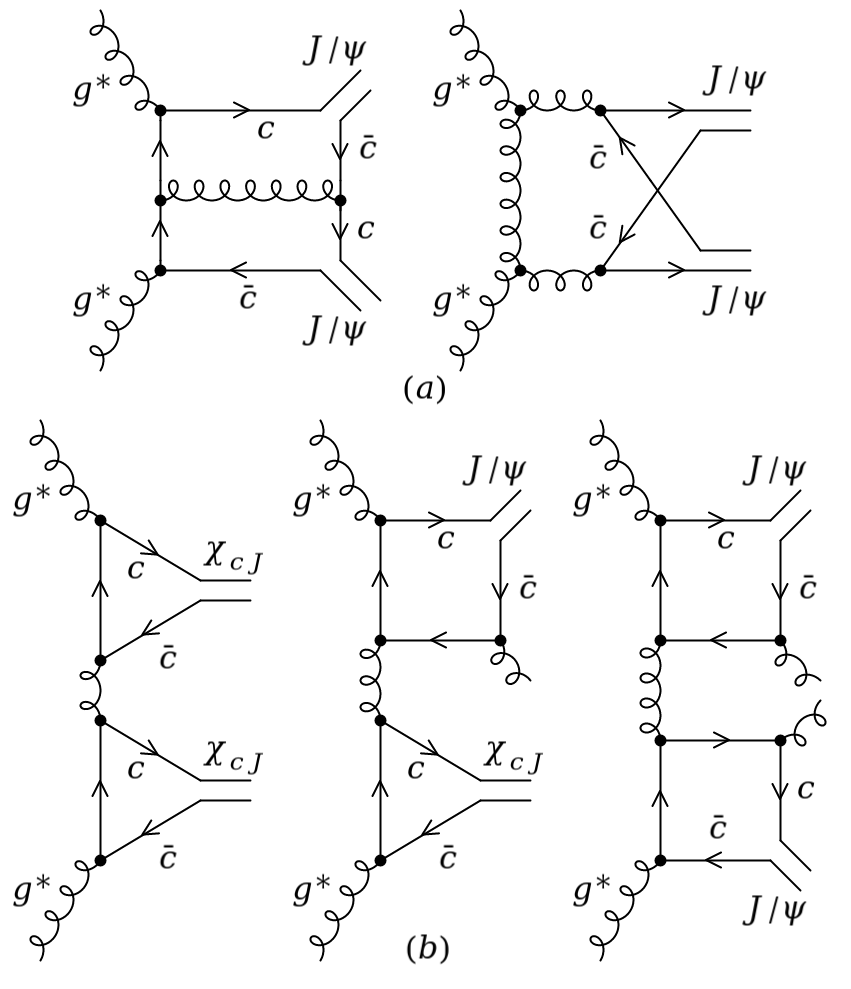}}\hfill

\caption{
Examples of Feynman diagrams, contributing to the production of: a) $J/\psi$ pair, subprocess (\ref{CS-1}); b) $J/\psi$ and $\chi_{cJ}$ pairs via CS mechanism, subprocesses (\ref{CS-2-1}) --- (\ref{CS-2}).}
\label{fig:CS}
 \end{center}
\end{figure}

To preserve consistency with our previous studies \cite{21,26}, here we employ the 
$k_T$-factorization approach \cite{27,28,29,30}.
This approach is based on the Balitsky-Fadin-Kuraev-Lipatov (BFKL) \cite{30,31} 
or Catani-Ciafaloni-Fiorani-Marchesini (CCFM) gluon evolution equations
and has certain technical advantages in the ease of including higher-order pQCD 
radiative corrections (namely, the leading-logarithm part of NLO + NNLO + ...  terms 
corresponding to real gluon emissions) in the form of transverse momentum dependent 
(TMD, or unintegrated) gluon density in a proton. It can be used as a convenient 
alternative to explicit higher-order pQCD calculations. A detailed description of this approach can be found, for example, in review \cite{32}.

First, we consider the $\mathcal{O}(\alpha^4_s)$ off-shell gluon-gluon fusion subprocess
(where the initial gluons have nonzero transverse momenta and are then off-shell)
which represents the leading order CS contribution:
\begin{gather}
  g^* + g^* \to c\bar c \left[^3S_1^{(1)}\right] + c\bar c \left[^3S_1^{(1)}\right].
  \label{CS-1}
\end{gather}
\noindent
Additionally, we take into account some subleading subprocesses:
\begin{gather}
g^* + g^* \to c\bar c \left[^3P_J^{(1)}\right] + c\bar c \left[^3P_J^{(1)}\right],
\label{CS-2-1}
\end{gather}
\begin{gather}
g^* + g^* \to c\bar c \left[^3P_J^{(1)}\right] + c\bar c \left[^3S_1^{(1)}\right] + g,
\label{CS-2-2}
\end{gather}
\begin{gather}
g^* + g^* \to c\bar c \left[^3S_1^{(1)}\right] + c\bar c \left[^3S_1^{(1)}\right] + g + g,
 \label{CS-2}
\end{gather}
\noindent
together with the feeddown contributions from excited states $\psi^\prime$ decaying 
into $J/\psi$ mesons. These subprocesses are formally suppressed by extra powers
of $\alpha_s$, or smaller values of $P$-wave mesonic wave functions, 
or $\chi_{c}\to J/\psi+\gamma$ decay branchings, but have an important kinematic 
property of filling the region of large invariant masses and large rapidity separation
in $J/\psi$ pairs. The typical Feynman diagrams are presented in Fig.~\ref{fig:CS}.

The production amplitudes of (\ref{CS-1}) --- (\ref{CS-2}) contain spin and color 
projection operators \cite{33,34,35,36} which guarantee the proper quantum numbers 
of final state charmonia. 
In accordance with the $k_T$-factorization prescription, initial gluons have non-zero
transverse four-momenta $k^2_{1T} = - \mathbf{k}^2_{1T} \neq 0$ and 
$k^2_{2T} = - \mathbf{k}^2_{2T} \neq 0$, and their polarization vectors have an 
admixture of longitudinal component. The summation over polarizations is carried out
with $\sum \epsilon^{\mu}\epsilon^{*\nu} = \mathbf{k}^{\mu}_T\mathbf{k}^{\nu}_T/\mathbf{k}^2_T$ \cite{27,28,29,30}. 
This expression converges to the ordinary $g^{\mu\nu}$ in the collinear limit $\mathbf{k}_T \to 0$ 
after averaging over the azimuthal angle. 
The gauge invariant expressions for all these amplitudes have been obtained 
earlier\cite{37} and implemented into the Monte-Carlo event generator 
\textsc{pegasus}\cite{38}.

Now we turn to another class of processes which constitute
the key topic of our study, namely, the fragmentation contributions. 
The fragmentation approach is known to be valid at high transverse momenta 
$p_T \gg m_\psi$. 
The relevant fragmentation functions (FF) $\mathcal{D}^{\mathcal{H}}_{a}(z,\mu^2)$ 
describing the transition of parton $a$ into charmonium state $\mathcal{H}$ 
through a number of intermediate $Q\bar{Q}$ states can be presented as a series:
\begin{gather}
\mathcal{D}^{\mathcal{H}}_{a}(z,\mu^2) = \sum_{n} d^{n}_{a}(z,\mu^2)\langle\mathcal{O}^{\mathcal{H}}[n]\rangle
\end{gather}
\noindent
where $n$ labels the intermediate (CS or CO) state, and 
$\langle{O}^{\mathcal{H}}[n]\rangle$ are the corresponding LDMEs.
In this way, the single charmonia production cross section can be written as
\begin{gather}
\begin{split}
\frac{d\sigma (pp \to \mathcal{H} + X)}{dp_T} = \int \frac{d\sigma(pp \to g^*)}{dp^{(g^*)}_T}\mathcal{D}^{\mathcal{H}}_g(z,\mu^2)\delta (z-p/p^{(g)})dz\\ + \int \frac{d\sigma(pp \to c\bar{c})}{dp^{(c)}_T}\mathcal{D}^{\mathcal{H}}_c(z,\mu^2)\delta (z-p/p^{(c)})dz,
\end{split}
\label{xsection-fragm}
\end{gather}
where $p^{(g^*)}$, $p^{(c)}$ and $p$ are the momenta of the gluon, charmed quark and outgoing
charmonium state $\mathcal{H}$, respectively. We take into consideration the channels
$g \to c\bar{c} [^3S^{(8)}_1]$, $g \to c\bar{c}[^3P^{(8)}_J] + g$ and
$c \to c\bar{c}[^3S^{(1)}_1] + c$ 
giving sizeable contributions to the $S$-wave charmonia ($J/\psi$ and $\psi'$ mesons),
$g \to c\bar{c}[^3P^{(1)}_J] + g$, $g \to c\bar{c}[^3S^{(8)}_1]$ and
$c \to c\bar{c}[^3P^{(1)}_J] + c$ 
contributing to the $P$-wave charmonia ($\chi_{cJ}$ mesons with $J = 1, 2$). The 
$\chi_{c0}$ channel is neglected because of low branching fraction to $J/\psi$.
The present list is more complete in comparison with our previous paper \cite{21} 
where we refer solely to the $g \to c\bar{c}[^3S^{(8)}_1]$ channel.

 \begin{figure}
\begin{center}
{\includegraphics[width=.8\textwidth]{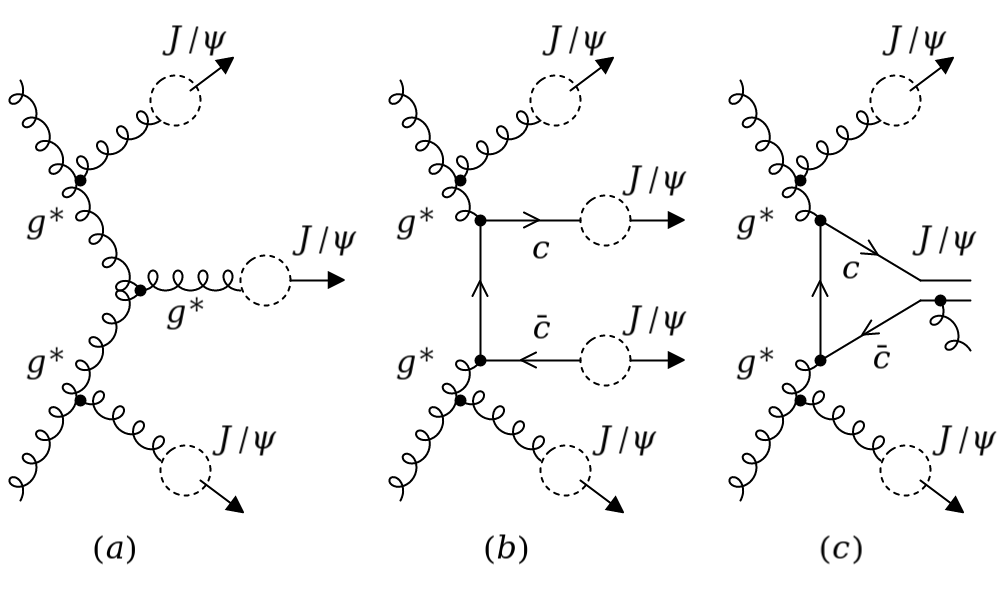}}\hfill

\caption{
Fragmentation contributions to the $J/\psi$ pair production that takes into account multiple gluon emissions 
for subprocesses: a) $g^* + g^* \to g^*$, b) $g^* + g^* \to c + \bar{c}$, c) $g^* + g^* \to c\bar{c}[^{2S+1}L^{(a)}_J]$. 
Circles represent the transition of parton into $J/\psi$ meson via fragmentation mechanism.
}
\label{fig:fragm}
 \end{center}
\end{figure}

The key point of our consideration is that any hard subprocess (giving or not giving
rise to a $J/\psi$ meson) is always accompanied by a number of gluons radiated during 
the non-collinear QCD evolution, and these gluons can fragment into additional $J/\psi$
particles.
We calculate the corresponding contributions by collecting all possible parton 
fragmentation combinations. At high energies, the QCD evolution of gluon cascade can
be described by the CCFM equation. This equation smoothly interpolates between the 
small-$x$ BFKL gluon dynamics and high-$x$ DGLAP one, and, therefore, 
provides us with a suitable tool for our phenomenological study. 
The numerical calculations split in two steps. 
First, we simulate the perturbative production of gluons, quarks, and charm pairs in 
the corresponding off-shell gluon-gluon fusion subprocesses and then reconstruct the 
CCFM gluon evolution ladder using the \textsc{cascade} Monte Carlo generator \cite{39}.
After that, one can collect $J/\psi$ pairs by looking over all possible combinations 
of mesons originating from charmed pairs, charmed quarks and gluons 
(including those formed in the evolution cascade). The combinatorics is rather large, 
because any fragmenting parton can be paired with any other parton.

For the initial hard subprocesses, we consider
\begin{gather}
\begin{split}
g^* + g^* \to g^*, \quad g^* + g^* \to c + \bar{c}, \quad g^* + g^* \to q + \bar{q}, \\ g^* + g^* \to c\bar{c}\left[\ ^3P^{(1,8)}_J\right], \quad g^* + g^* \to c\bar{c}\left[\ ^3S^{(1)}_1\right] + g,
\end{split}
\label{subprocesses-fragm}
\end{gather}
\noindent
where $q$ denotes light quarks. 
The typical diagrams of subprocesses~(\ref{subprocesses-fragm}) with reconstructed gluon
ladders and possible channels of fragmentation are presented in Fig.~\ref{fig:fragm}. 
The CO channel $g^* + g^* \to c\bar{c}[^3S^{(8)}_1]$ is excluded from subprocesses 
(\ref{subprocesses-fragm}) in order to avoid double counting with the fragmentation 
mechanism $g^* + g^* \to g^* \to c\bar{c}[^3S^{(8)}_1]$. 
By the way, we have found that these two subprocesses perfectly match one another
in the ATLAS kinematic range.
 
The expressions for fragmentation functions at the initial scale $\mu^2_0 = m^2_{\psi}$ 
can be found \cite{40}. For the fragmentation $g\to c\bar{c}[^3P^{(1,8)}_J] + g$ 
we use the expression derived very recently \cite{41}, with the mass of the emitted 
gluon (considered as regularization parameter) $m_g = m_c =1.5$ GeV.
{This expression is positive-definite, smooth and vanishes at the endpoints
$z=0$ and $z=1$.}
The shapes of fragmentation functions are modified by the final state gluon radiation;
these effects can be described in proper way with the DGLAP evolution equation:
\begin{gather}
 \frac{d}{d\ln\mu^2}\left(\begin{array}{lr} \mathcal{D}^{\mathcal{H}}_c \\ \mathcal{D}^{\mathcal{H}}_g \end{array} \right)=\frac{\alpha_s(\mu^2)}{2\pi}\left(\begin{array}{lr} P_{cc} & P_{gc} \\ P_{cg} & P_{gg} \end{array}\right)\otimes\left(\begin{array}{lr} \mathcal{D}^{\mathcal{H}}_c \\ \mathcal{D}^{\mathcal{H}}_g \end{array}\right),
\label{DGLAP}
\end{gather}
where $P_{ab}$ are the usual LO DGLAP splitting functions. 
According to the non-relativistic QCD approximation, we set the charmed quark mass 
to $m_c = m_{\psi}/2$ and then solve the DGLAP equation (\ref{DGLAP}) numerically with 
the proper LDME's.

\begin{figure}
\begin{center}
{\includegraphics[width=.5\textwidth]{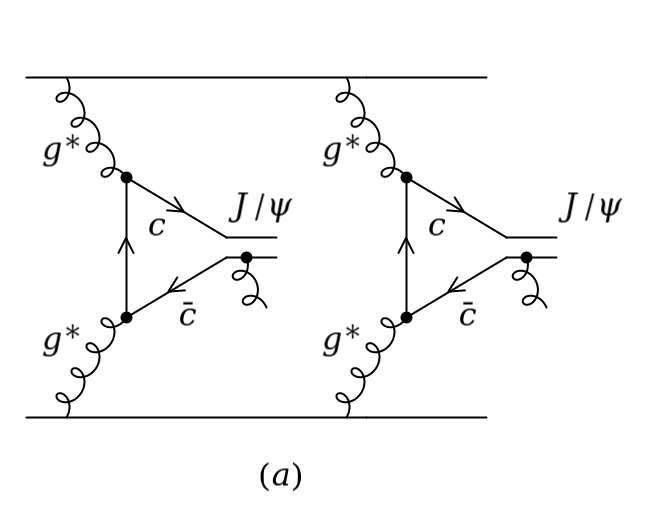}}\hfill
{\includegraphics[width=.5\textwidth]{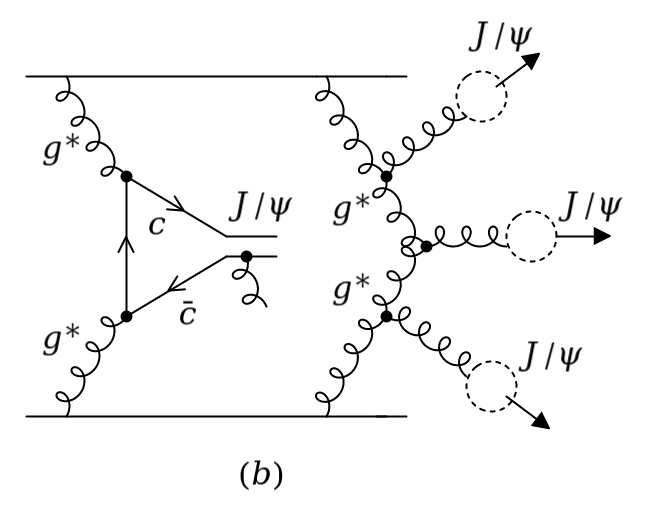}}\hfill
\caption{
Example of double $J/\psi$ production for double parton scattering in conventional scheme of calculation (a) and modified scheme with multiple gluon radiation (b). Circles represent the transition of parton into $J/\psi$ meson via fragmentation mechanism.
}
\label{fig:DPS_scheme}
 \end{center}
\end{figure}

The last contribution taken into consideration refers to the double parton scattering 
(DPS) mechanism. According to the standard factorization formula \cite{42,43}, this 
contribution can be presented in a simple form:
\begin{gather}
\sigma_{\rm DPS}(pp \to J/\psi J/\psi + X) = \frac{1}{2}\frac{\sigma_1(pp \to J/\psi + X)\cdot\sigma_2(pp \to J/\psi + X)}{\sigma_{\rm eff}},
\label{DPS}
\end{gather}
where the factor $1/2$ prevents double counting between identical particles. 
The "effective cross section" $\sigma_{\rm eff}$ is a normalization factor which
encodes all "DPS unknowns" into a single parameter and represents the effective 
transverse overlap of partonic interactions that constitute the DPS process. 
In general, it can be regarded as free parameter which should be extracted from the data.
The decomposition (\ref{DPS}) of the DPS cross section into two individual single parton
scattering (SPS) factors without correlation and interference between them is acceptable
for ATLAS kinematics region.
The inclusive cross sections $\sigma(pp \to J/\psi + X)$ involved in~(\ref{DPS})
are calculated in the $k_T$-factorization approach supplemented by the NRQCD formalism 
in a standard way (see, for example, \cite{44} and references therein). 
When calculating the DPS cross section, we also take into account all the accompanying
fragmentation contributions
(including all possible combinations of radiated partons, see Fig.~\ref{fig:DPS_scheme}).


In the numerical calculations we use TMD gluon densities in a 
proton obtained from a numerical solution of CCFM evolution equation, 
namely, JH'2013 set 1 and JH'2013 set 2 \cite{45}. 
The input parameters of JH'2013 set 1 gluon distribution have been fitted to the proton 
structure function $F_2(x,Q^2)$, 
whereas the input parameters of JH'2013 set 2 gluon were fitted to the both structure 
functions $F_2(x,Q^2)$ and $F^c_2(x,Q^2)$. According to \cite{45}, we 
use the two-loop formula for the QCD coupling $\alpha_{s}$ with $n_f = $ 4 active quark 
flavors and $\Lambda_{\rm QCD} = $ 200 MeV.

The charmonia LDMEs have been defined earlier \cite{44} from a global fit to the LHC 
data. Using these LDMEs, we reproduce all of the available data on charmonia production 
at the LHC conditions.
A comprehensive information on the fitting procedure can be found in \cite{44}. 
The masses and the branching fractions of all particles involved into calculations 
are taken from \cite{46}. 
The DPS effective cross section is chosen as  $\sigma_{\rm eff} = 13.8$~mb, which was 
extracted from a fit to the latest LHCb data on the double $J/\psi$ production
(see \cite{21} for more details). This value is very close to the generally accepted 
value $\sigma_{\rm eff} = 15$~mb\cite{47}.

\begin{figure}
\begin{center}
{\includegraphics[width=.5\textwidth]{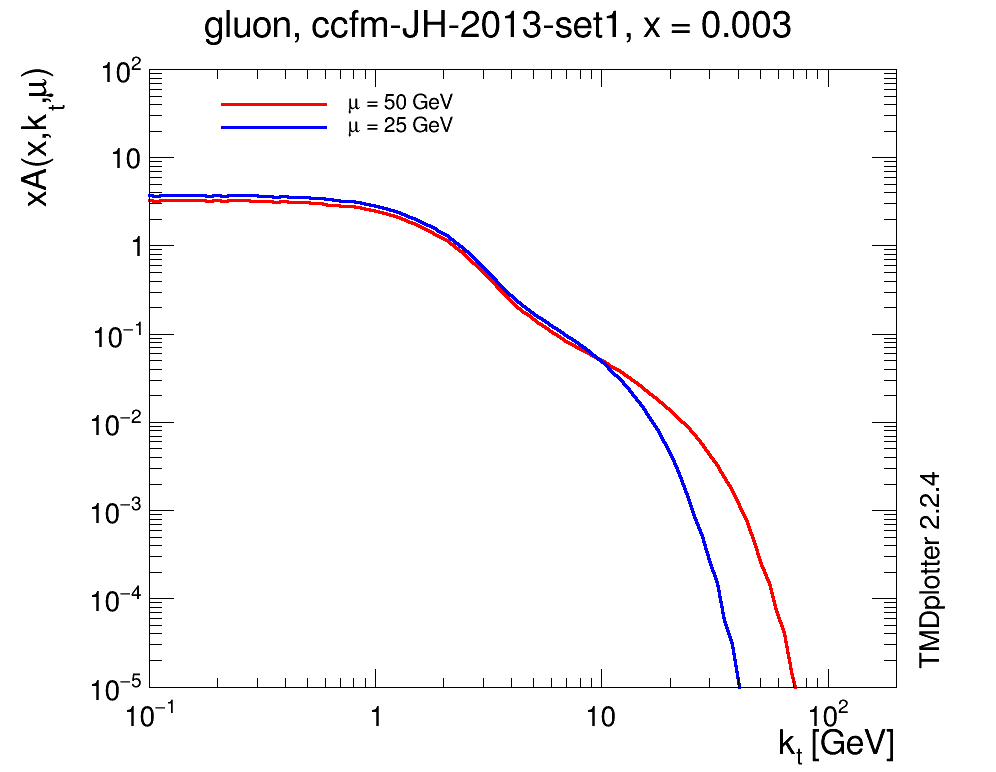}}\hfill
{\includegraphics[width=.5\textwidth]{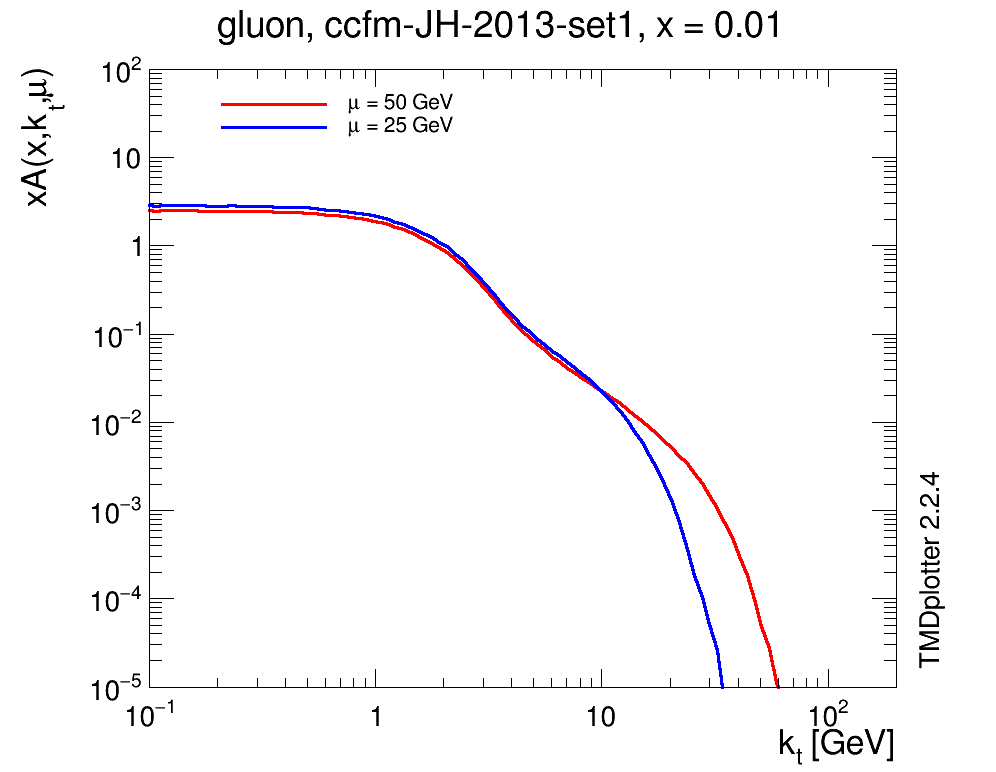}}\hfill

\caption{The effect of scale variations in TMDs. By changing the factorization scale,
 we jump between the red and blue lines; in the region $\mu_F\simeq k_T$ the
 difference may be as large as several orders of magnitude.}
\label{fig:TMD}
 \end{center}
\end{figure}

\begin{figure}
\begin{center}
{\includegraphics[width=.48\textwidth]{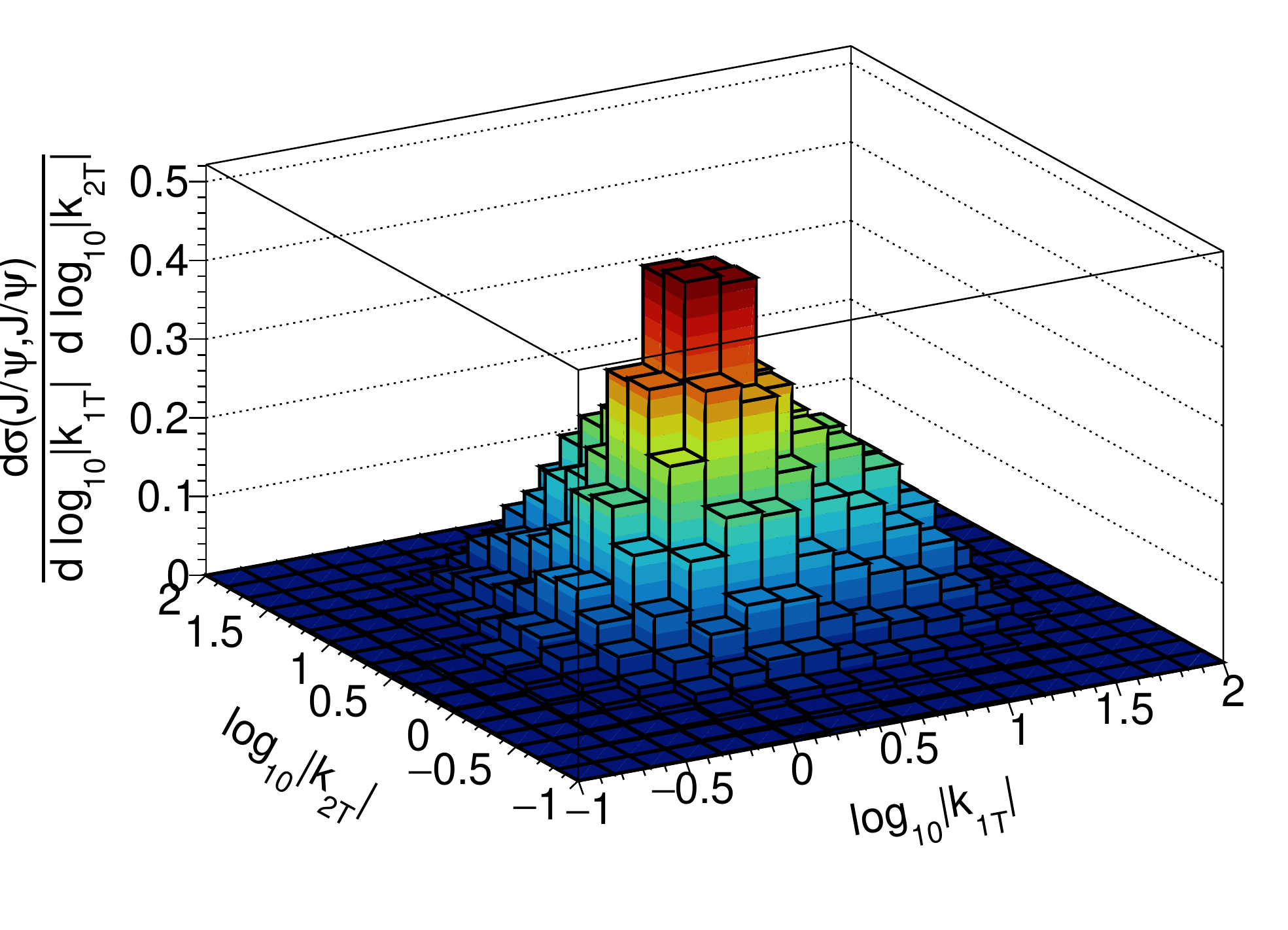}}\hfill
{\includegraphics[width=.48\textwidth]{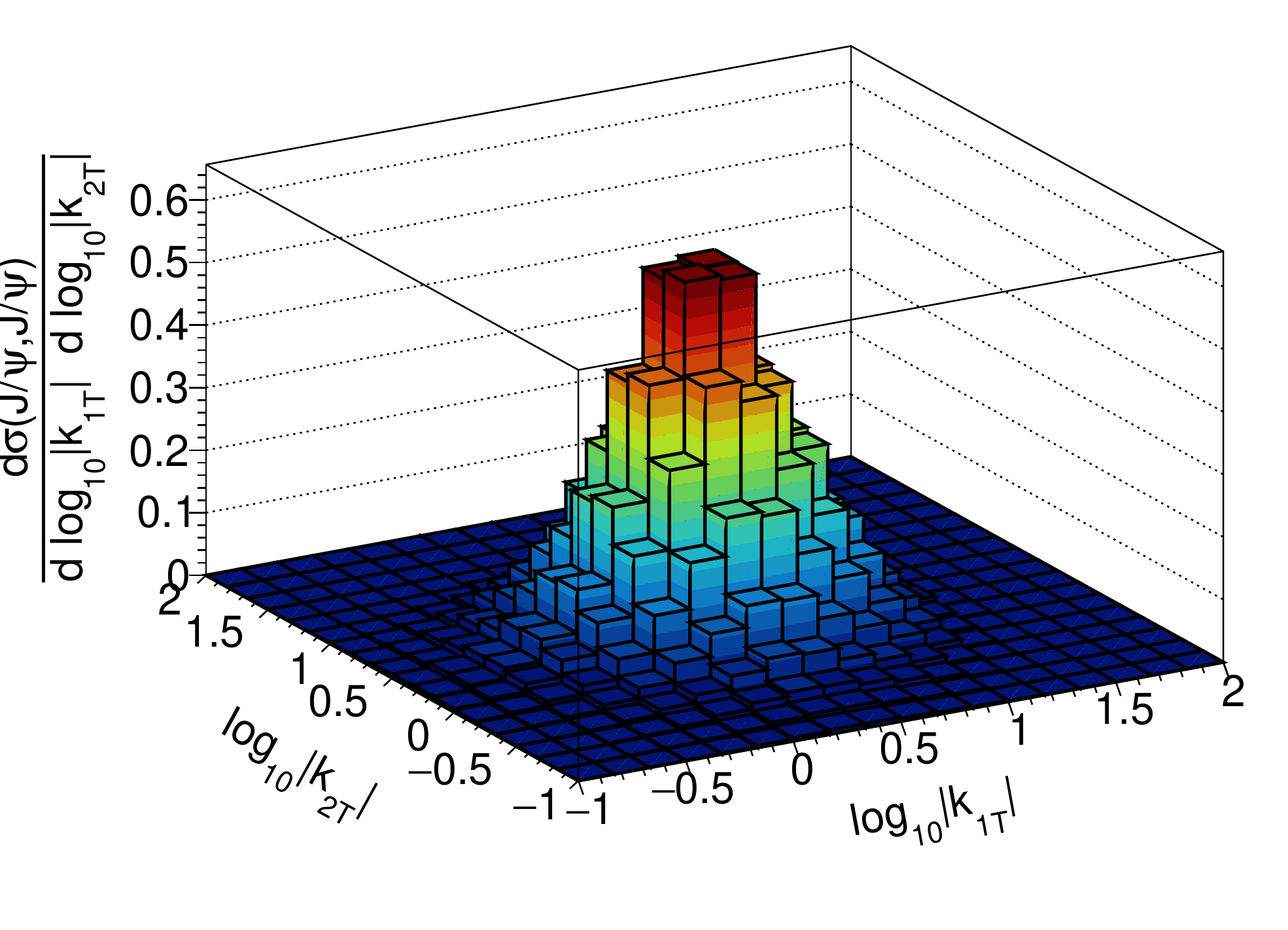}}\hfill
{\includegraphics[width=.48\textwidth]{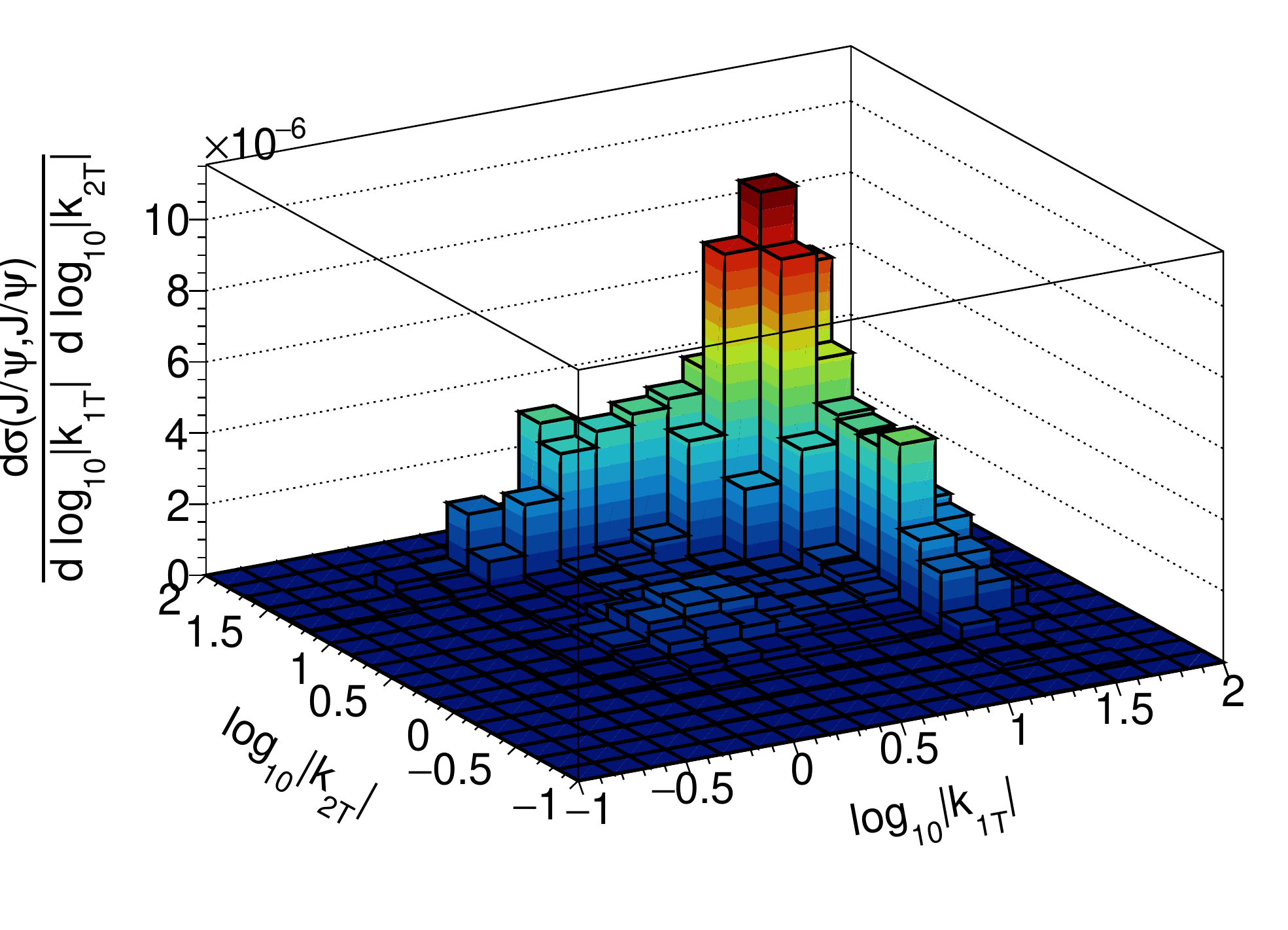}}\hfill
{\includegraphics[width=.48\textwidth]{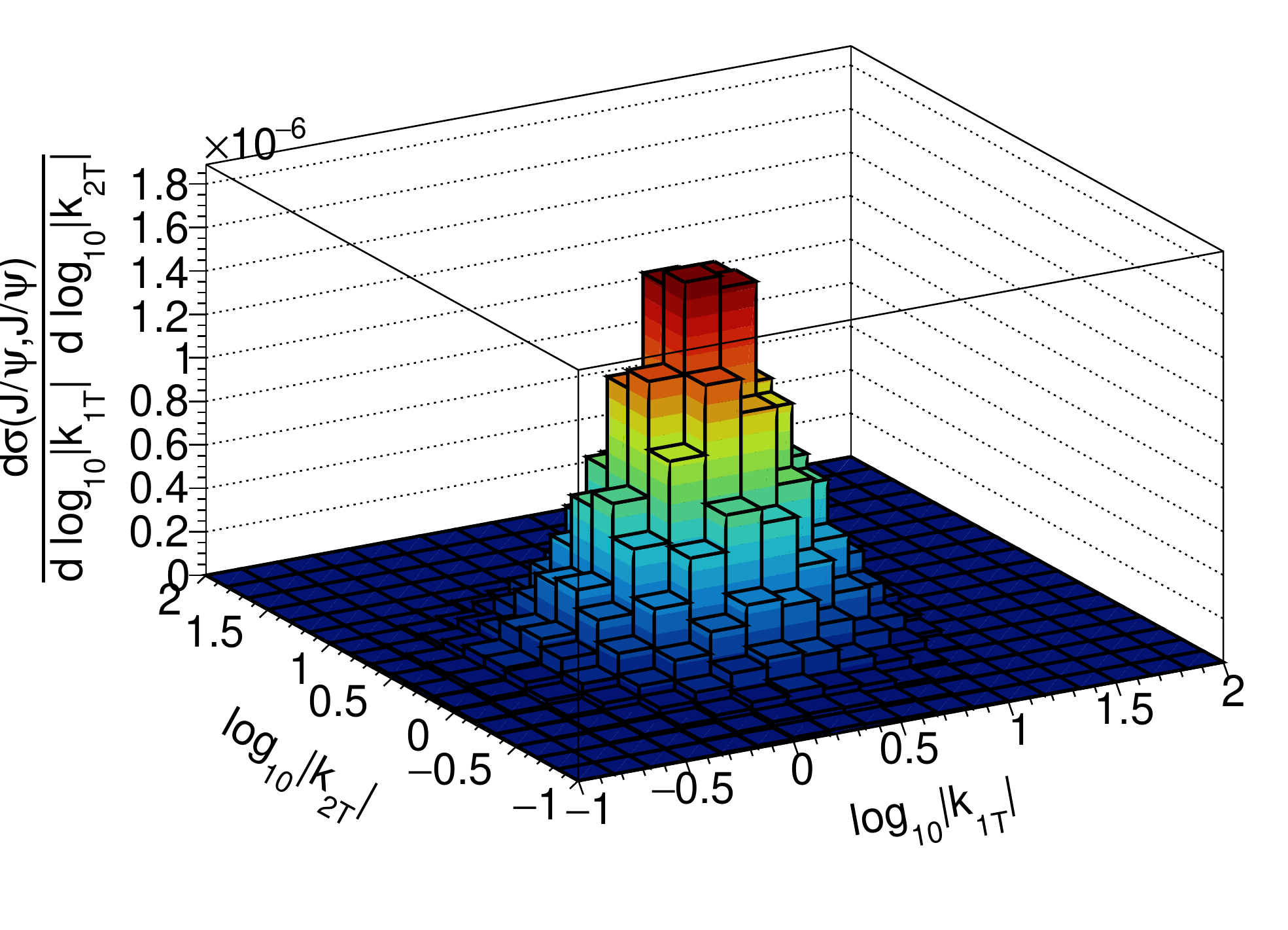}}\hfill
{\includegraphics[width=.48\textwidth]{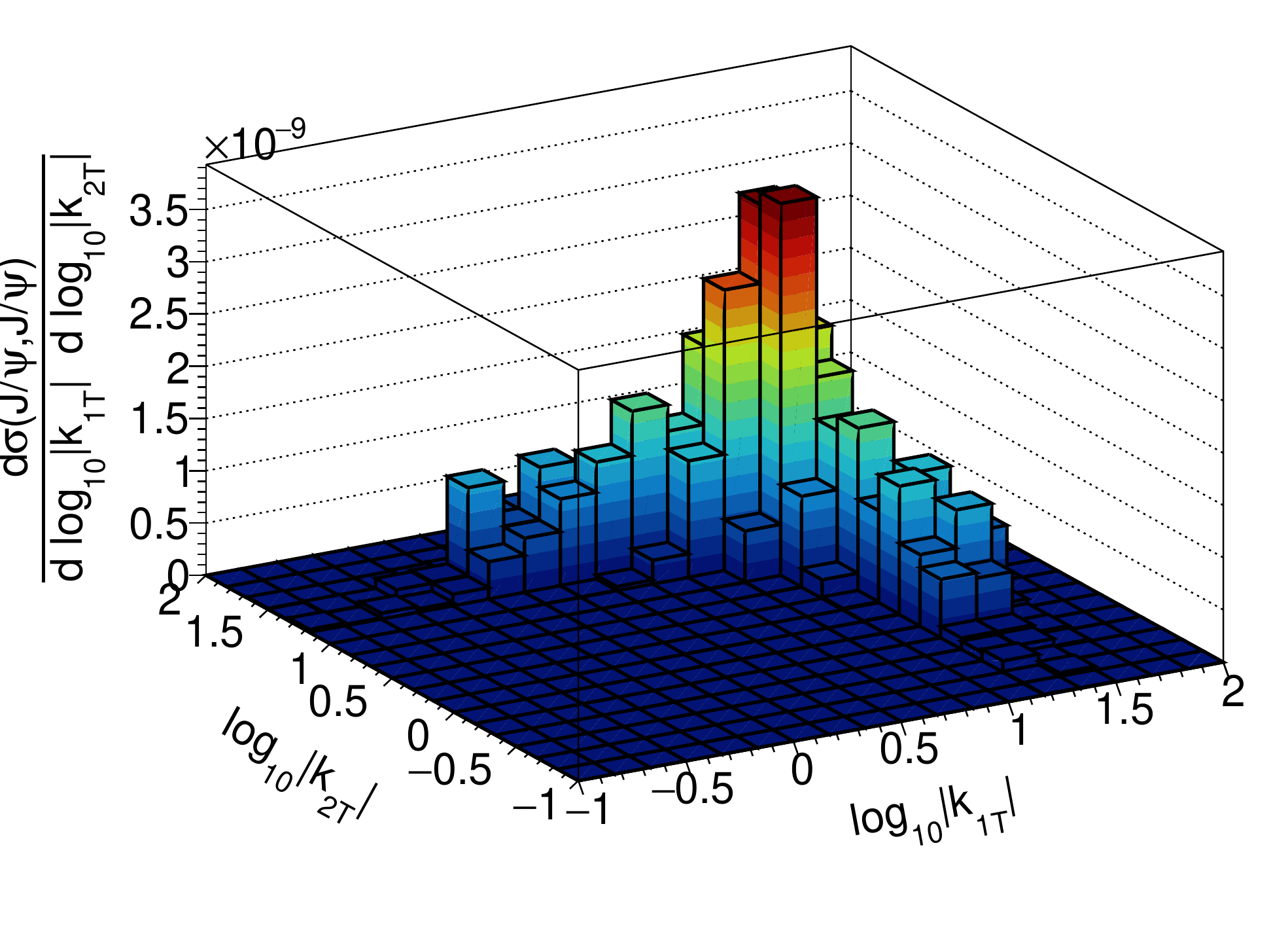}}\hfill
{\includegraphics[width=.48\textwidth]{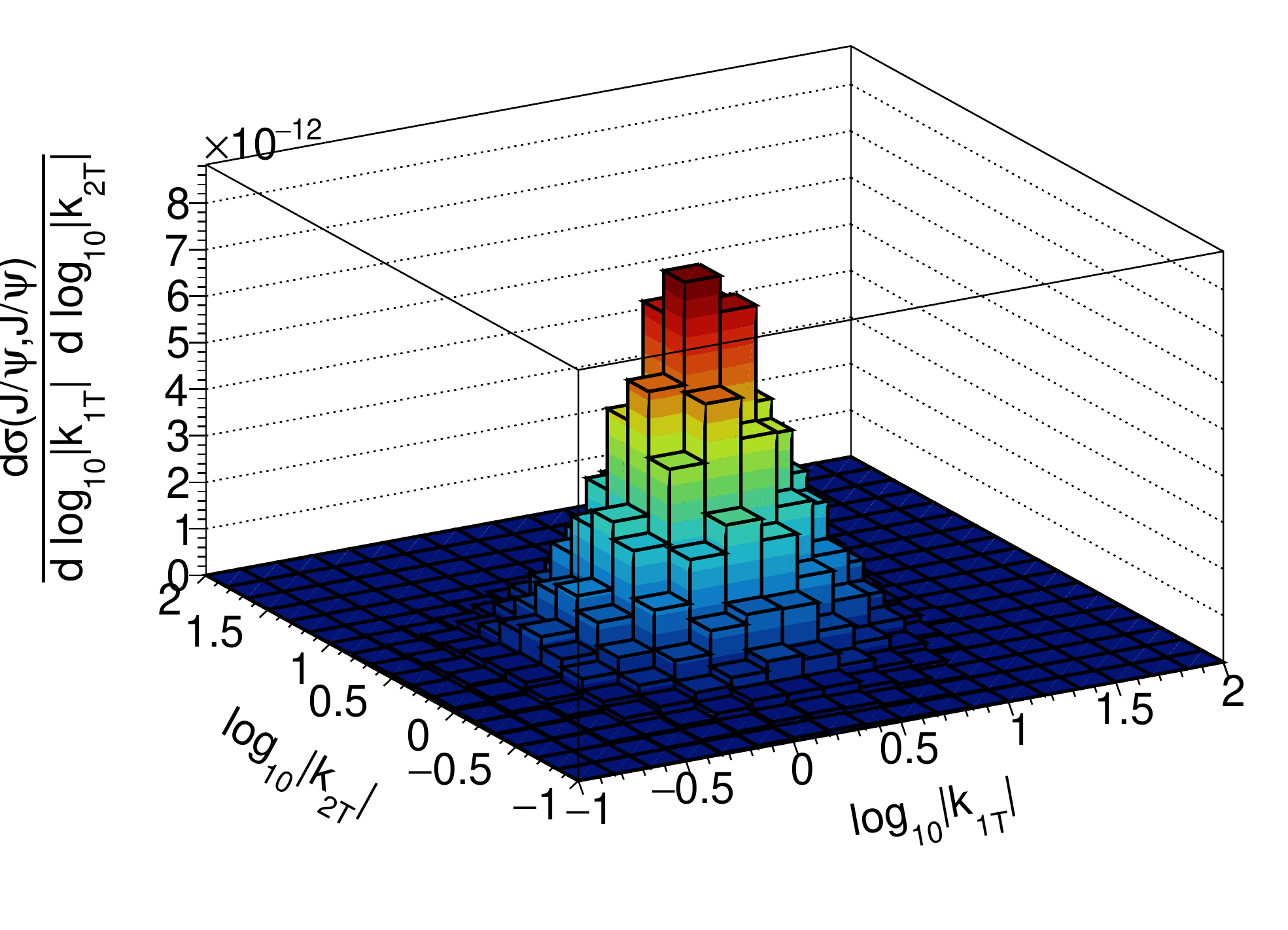}}\hfill

\caption{
Double differential distributions in the gluon transverse momenta, $d\sigma/d\log_{10}{|{\mathbf k}_{1T}|} d\log_{10}{|{\mathbf k}_{2T}|}$; toy
calculations with differently chosen factorization scales and different toy matrix elements $|{\cal M}|^2$ for the partonic subprocess. 
Left column corresponds to $\mu^2_F = \hat{s} + \mathbf{Q}^2_T$, right column --- $\mu^2_F = \hat{s}/4$. 
Upper row, $|{\cal M}|^2\propto 1$; middle row, $|{\cal M}|^2\propto 1/\hat{s}^2$; 
lower row, $|{\cal M}|^2\propto 1/\hat{s}^4$. Calculations were performed with JH'2013 set 1.
}
\label{fig:scale_1}
 \end{center}
\end{figure}

\section{Choice of factorization scale} \indent

\begin{figure}
\begin{center}
{\includegraphics[width=.48\textwidth]{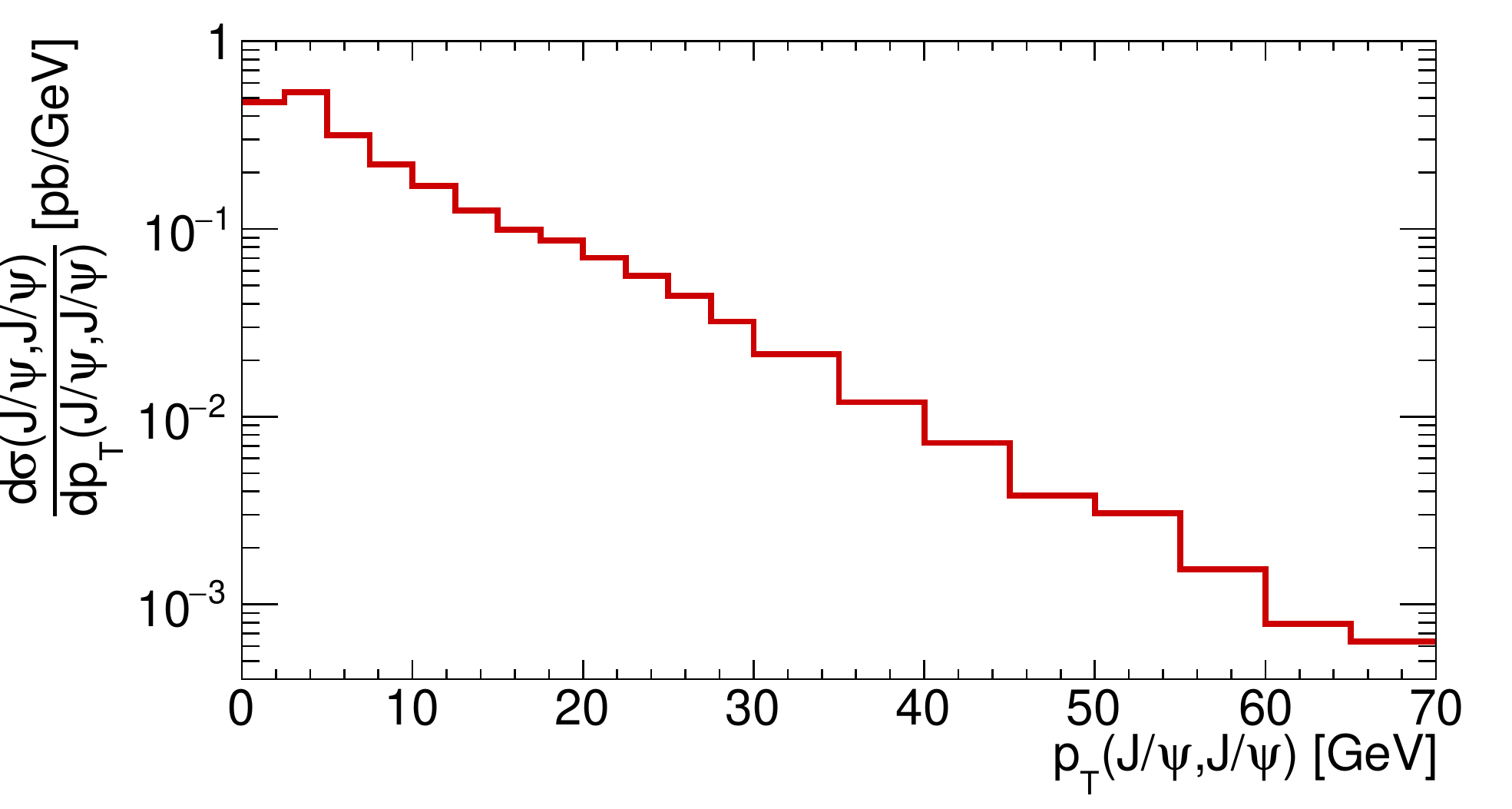}}\hfill
{\includegraphics[width=.48\textwidth]{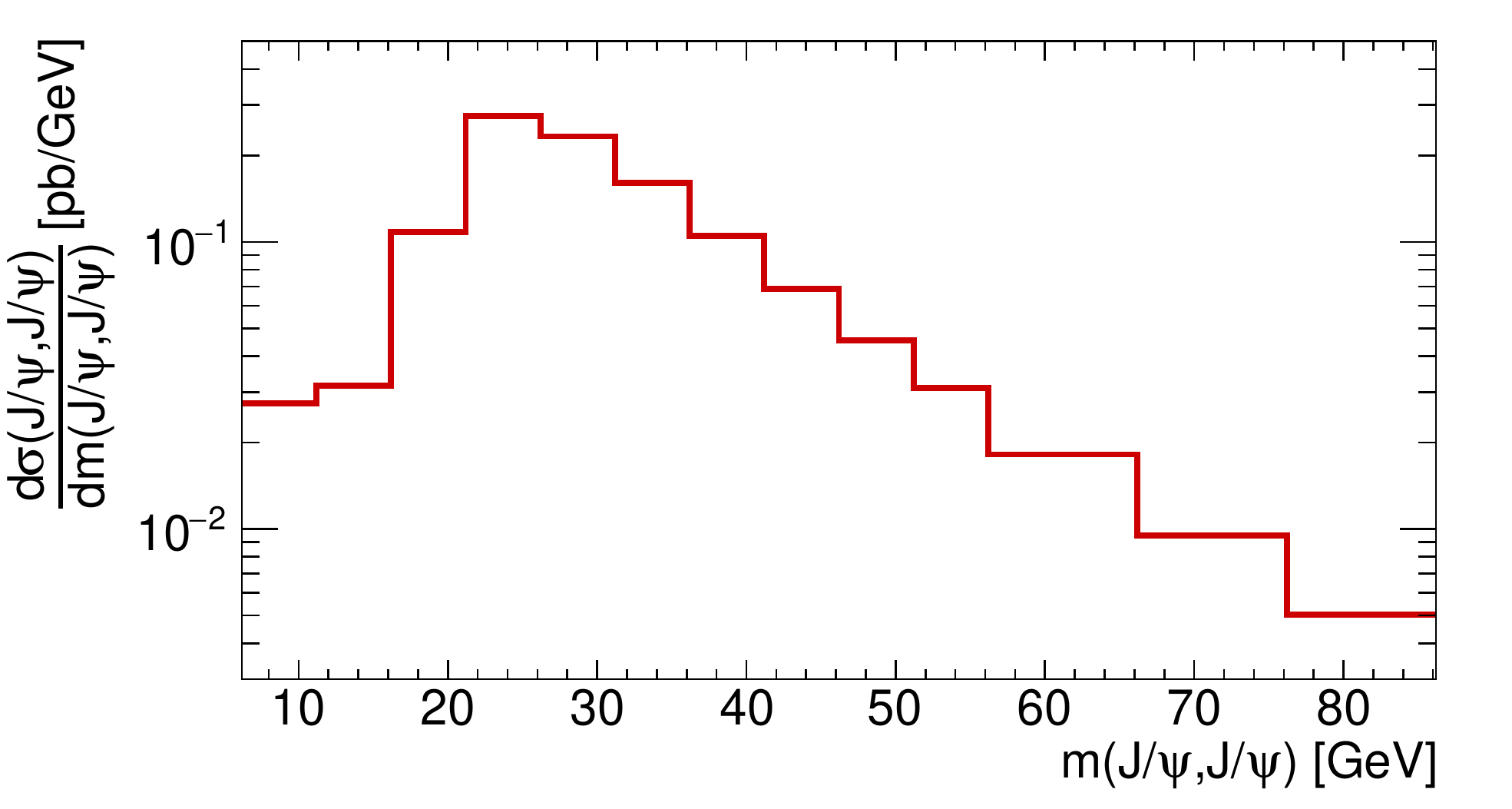}}\hfill
{\includegraphics[width=.48\textwidth]{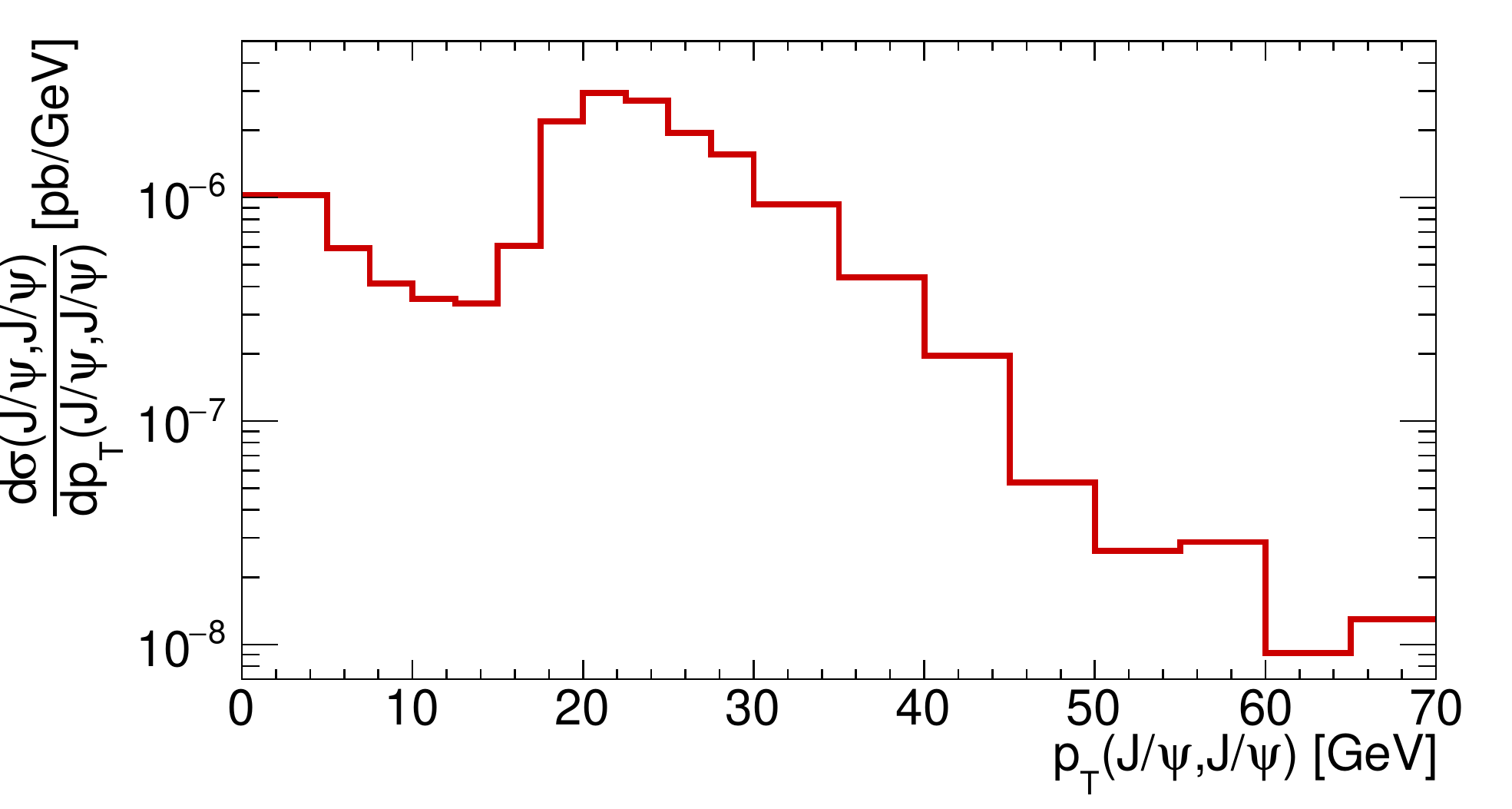}}\hfill
{\includegraphics[width=.48\textwidth]{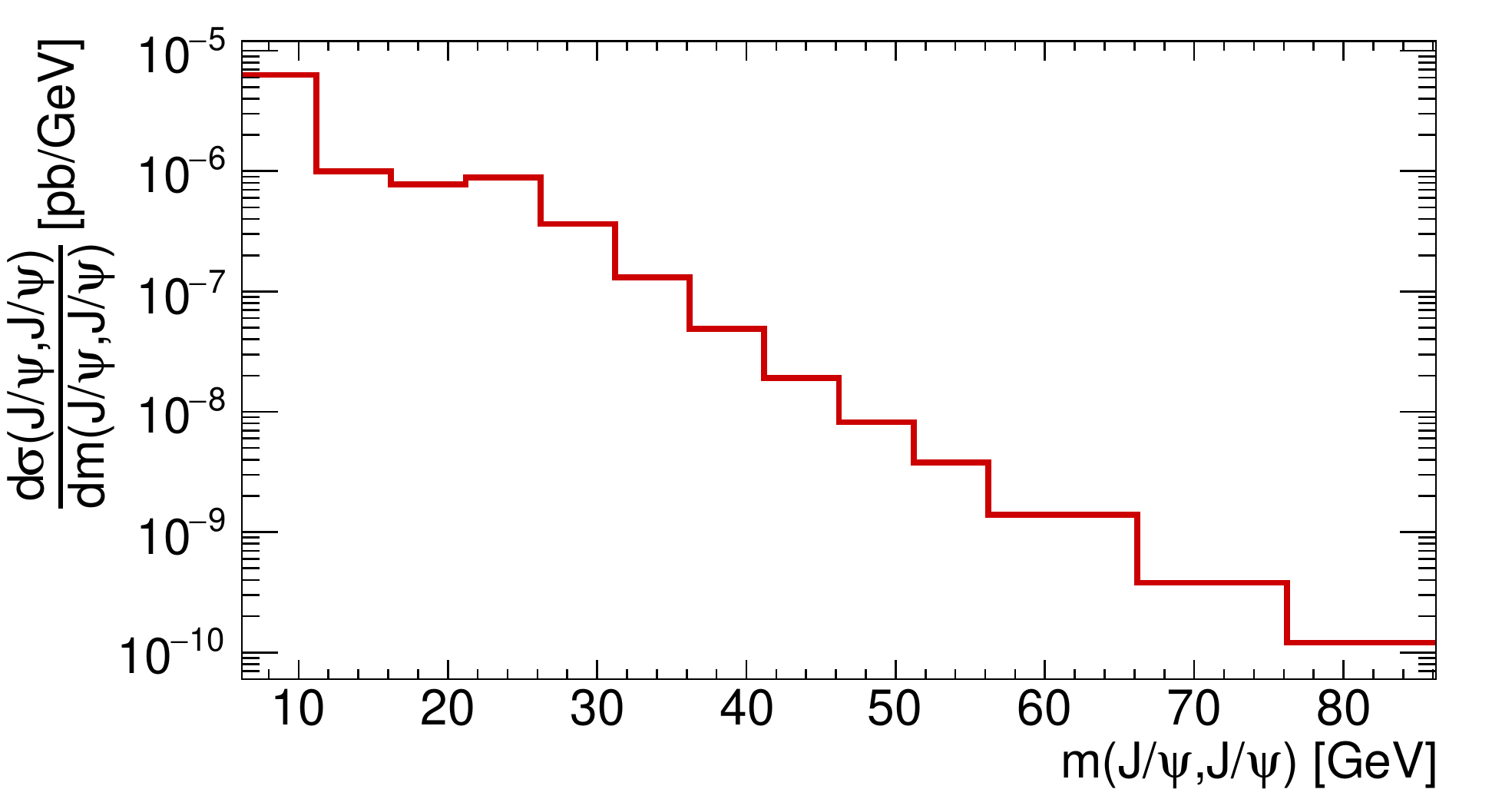}}\hfill
{\includegraphics[width=.48\textwidth]{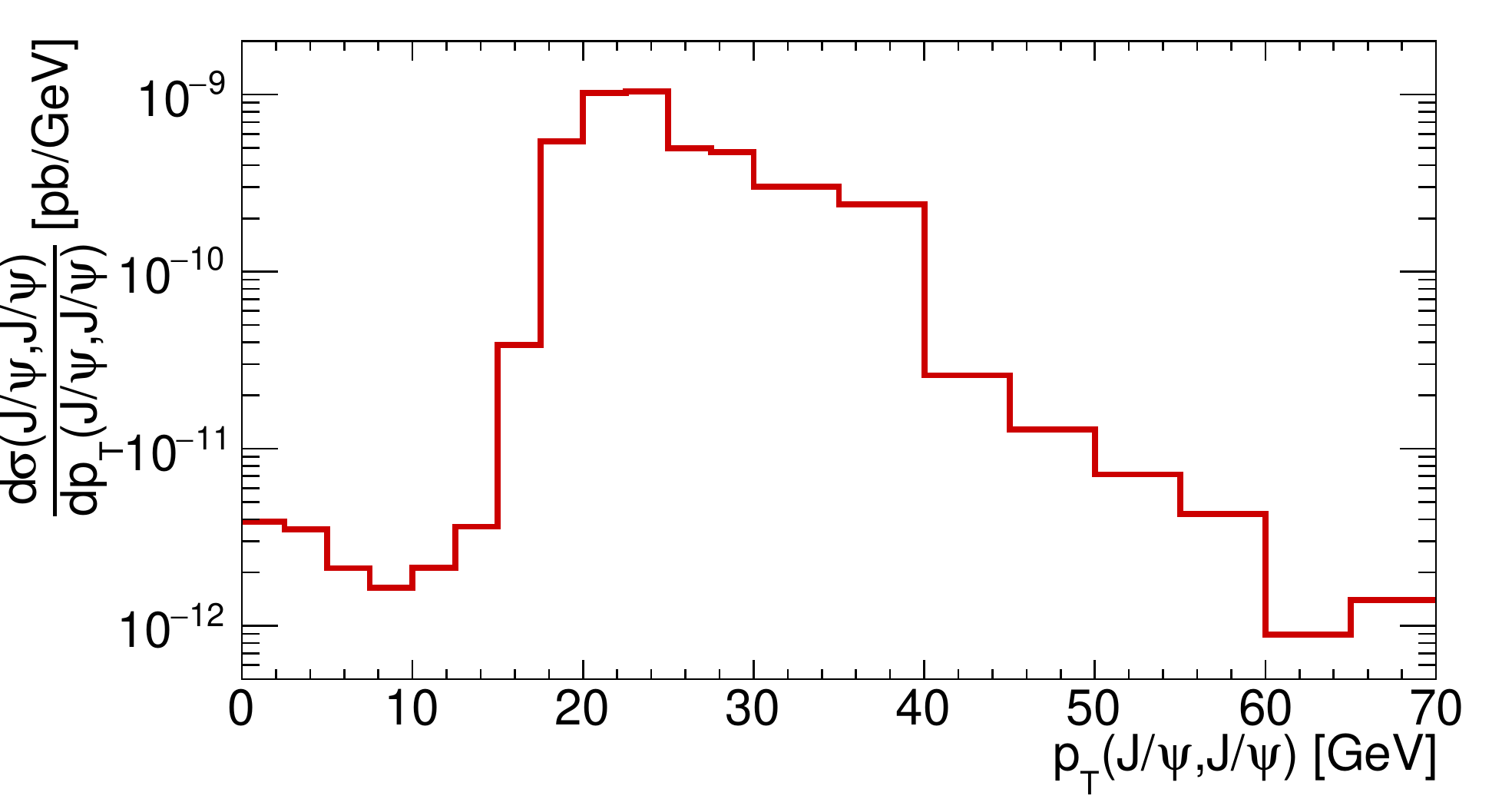}}\hfill
{\includegraphics[width=.48\textwidth]{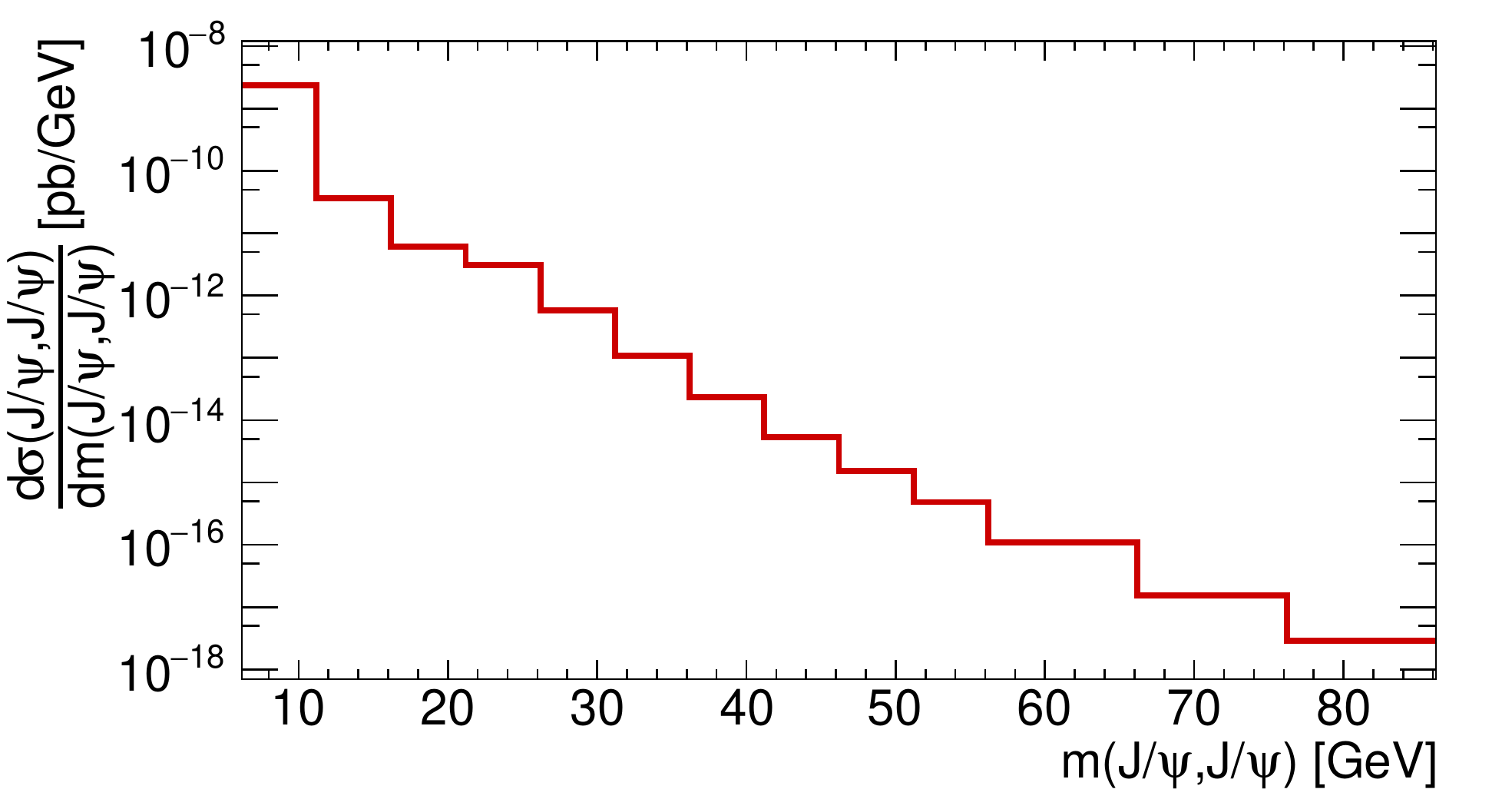}}\hfill
\caption{
The $J/\psi$ pair transverse momentum (left column) and the $J/\psi$ pair invariant mass (right column) distributions; 
toy calculations for $\mu^2_F = \hat{s} + \mathbf{Q}^2_T$ and different toy matrix elements for the partonic 
subprocess. Upper row, $|{\cal M}|^2\propto 1$; middle row, $|{\cal M}|^2\propto 1/\hat{s}^2$; lower row, $|{\cal M}|^2\propto 1/\hat{s}^4$. Calculations were performed with JH'2013 set 1.
}
\label{fig:scale_2}
\end{center}
\end{figure}

\begin{figure}
\begin{center}

{\includegraphics[width=.5\textwidth]{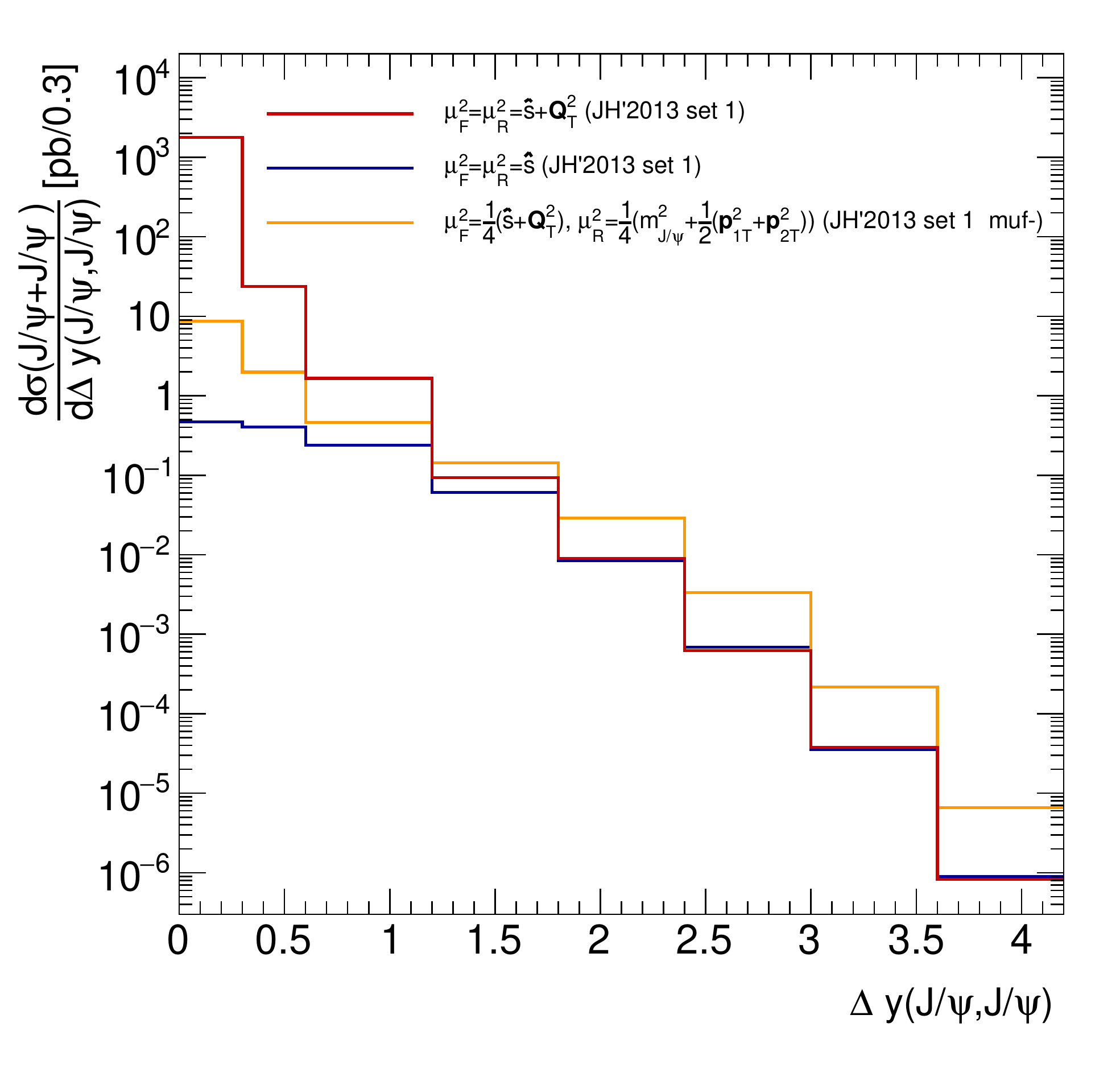}}\hfill
{\includegraphics[width=.5\textwidth]{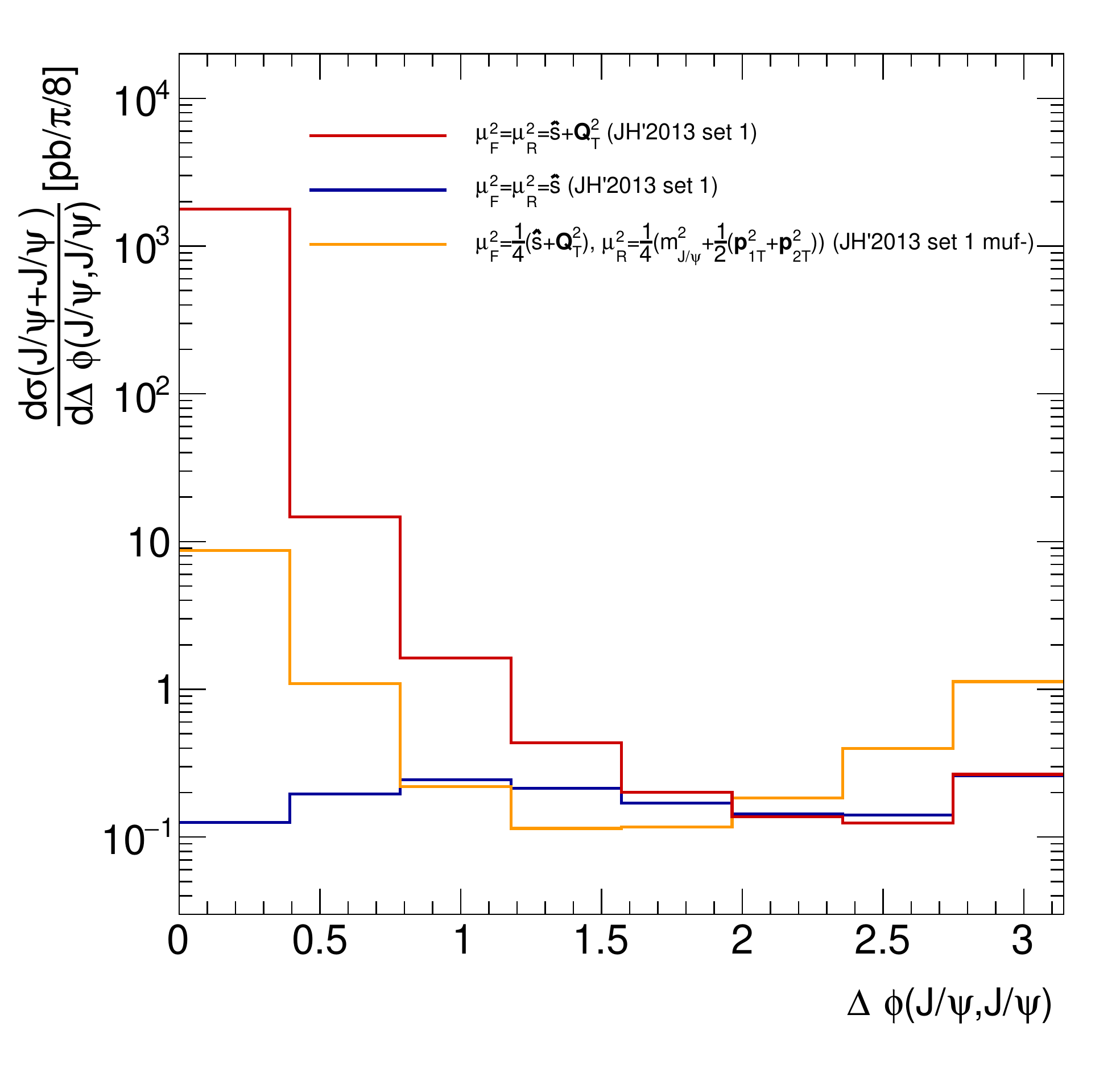}}\hfill
{\includegraphics[width=.5\textwidth]{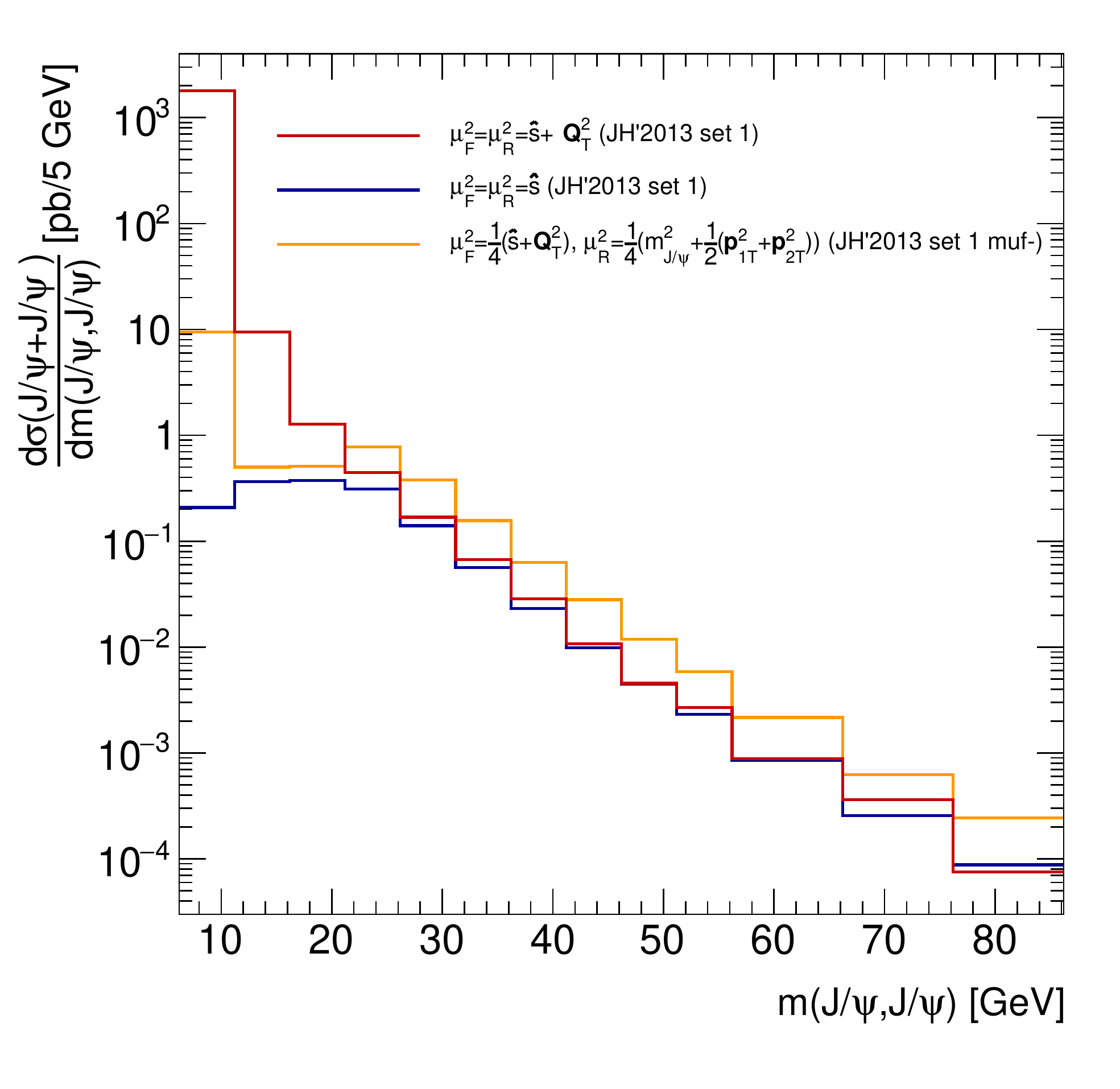}}\hfill
{\includegraphics[width=.5\textwidth]{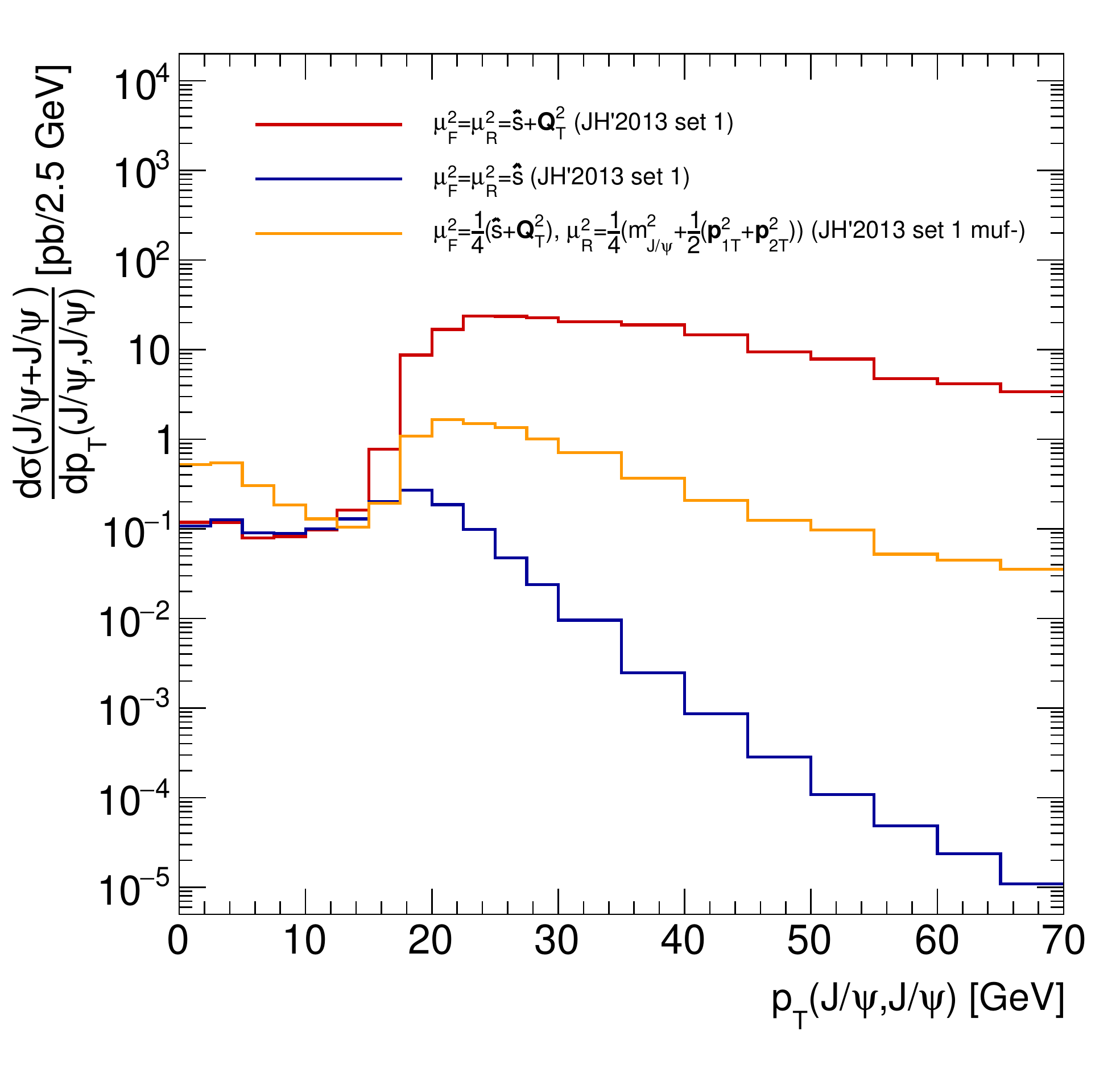}}\hfill
\caption{The effect of the choice of the factorization scale $\mu_F$ on the size and
the shape of the distributions in the 
 rapidity separation, $\Delta y(J/\psi,J/\psi)$ (left upper panel); 
 azimuthal angle difference, $\Delta \phi(J/\psi,J/\psi)$ (right upper panel); 
 $J/\psi$ pair invariant mass, $m(J/\psi,J/\psi)$ (left lower  panel); 
 $J/\psi$ pair transverse momentum, $p_T(J/\psi,J/\psi)$ (right lower panel).
The numerical instability is most pronounced at $p_T^2 \gg\hat{s}$, that is, 
when either $\hat{s}$ is small or $p_T$ is large.
}
\label{fig:scale_3}
\end{center}
\end{figure}

The choice of the factorization scale is a delicate, though very important issue.
In the CCFM evolution equation 
the factorization scale $\mu_F$ is defined as $\mu^2_F = \hat{s} + \mathbf{Q}^2_T$, 
where $\hat{s} = (k_1 + k_2)^2$ is the invariant energy of partonic subprocess and 
$\mathbf{Q}_T$ is the net transverse momentum of the initial gluon pair. It looks rather
natural to use the same definition of $\mu_F$ throughout all calculations, but this may 
cause problems in some cases. 

A peculiar property of the partonic subprocess~(\ref{CS-1}) 
is that its cross section drops sharply with increasing $\hat{s}$, so that $\hat{s}$
is typically not far from threshold (and, consequently, is small). At the same time, 
the $k_T$-spectra of the initial gluons are exponentially broad. This means that in an
arbitrary pair of gluons, the transverse momentum $|\mathbf{k}_T|$ of one of the gluons
is much larger than that of the other. As a result, the expression 
$\mu^2_F\propto \hat{s} + \mathbf{Q}^2_T$ reduces to $\mu^2_F \simeq \mathbf{k}^2_T$. 
However, such a definition is self-contradictory. Indeed, the gluon 
density $f_g(x,\mathbf{k}^2_T,\mu^2_F)$ must describe the probability distribution for 
$\mathbf{k}^2_T$ at any given $\mu_F$. Assume, we set some $\mu_F$ and generate a random 
gluon transverse momentum with the probability given by $f_g(x,\mathbf{k}^2_T,\mu^2_F)$. 
But then we come to a conflict with setting $\mu^2_F = \mathbf{k}^2_T$ which requires 
$\mu_F$ to be different from what was set originally. 

In fact, the condition $\mu^2_F\propto\mathbf{k}^2_T$ means that the full area of $\mu^2_F$ 
and $\mathbf{k}^2_T$ is not accessible, but we are only restricted to a one-dimensional 
trajectory tracing the functional dependence of $\mu^2_F$ on $\mathbf{k}^2_T$. Moreover,
this trajectory lies entirely on a steep slope, as the gluon densities steeply change 
around $\mathbf{k}^2_T = \mu^2_F$ from high values at $\mathbf{k}^2_T < \mu^2_F$ 
to low values at $\mathbf{k}^2_T > \mu^2_F$. This makes the calculation very unstable 
with respect to even small variations in the factorization scale, see Fig.\ref{fig:TMD}.

The above properties are further illustrated in Figs.~\ref{fig:scale_1} and~\ref{fig:scale_2}. 
Shown in Fig.~\ref{fig:scale_1} is the double differential distribution in the 
gluon transverse momenta 
$d\sigma/d\log_{10}{|\mathbf{k}_{1T}|}d\log_{10}{|\mathbf{k}_{2T}|}$ obtained 
from toy calculations with differently chosen factorization scales and with different 
toy matrix elements $|{\cal M}|^2$ for the partonic subprocess. Here we see that the 
choice $\mu^2_F = \hat{s}/4$ favors moderate gluon transverse momenta and the shape of
the distribution is insensitive to the properties of the matrix element. On the contrary,
the choice $\mu^2_F = \hat{s} + \mathbf{Q}^2_T$ makes the $\mathbf{k}^2_T$ values large
and highly unequal, and the $|\mathbf{k}_T|$ distribution is sensitive to the properties 
of $|{\cal M}|^2$.

The effect of numerical instability is shown in Fig.~\ref{fig:scale_2}. With sharper 
matrix elements, the leading role in $\mu_F$ transits from $\hat{s}$ to $\mathbf{Q}^2_T$ 
thus bringing us to an "unsafe" regime $\mu^2_F\simeq\mathbf{k}^2_T$.
It may be worth noting that the problem is rather general. Whatever the behavior 
of the matrix elements is, there always exists a kinematic region where $p_T^2 \gg\hat{s}$.
The case of $J/\psi$ pair production was only "lucky" to reveal the problem at
smaller energies and smaller $p_T$.

In Fig.~\ref{fig:scale_3} we extend our exercises from toy to real matrix elements and 
inspect the behavior of subprocess (\ref{CS-1}).
As is expected, the region of the biggest numerical sensitivity to the choice of $\mu_F$
is the region of the smallest invariant masses and, respectively, the region of the 
smallest rapidity difference $\Delta y(J/\psi, J/\psi)$.
The divergence between the different predictions is huge. At the same time, the curves
obtained with $\mu^2_F = (\hat{s} + \mathbf{Q}^2_T)/4$ are notably close to the 
experimental points as it is demonstrated at the Section 4. 
It is necessary to note that
these predictions were obtained with the TMD gluon densities JH'2013 set 1 muf$-$, which 
use the prefactor of $1/4$ in the definition of $\mu^2_F$ in the CCFM evolution and in the
fitting procedure.  

On the one hand,
the property that the fullrange gluonic phase space degenerates into a one-dimensional
line $\mu^2_F\propto\mathbf{k}^2_T$ can be regarded as pathological, 
though inevitable with increasing $p_T$. It leads to extraordinary 
sensitivity of the results to the choice of $\mu_F$. 
Since this kind of bad behavior is only seen at certain (rather special) kinematic 
conditions (subprocesses with very small $\hat{s}$), one can try 
to redefine the scale $\mu^2_F$ in this particular case. 
The definition of $\mu^2_F$ 
used in the TMD fitting procedure 
is related to the maximum angle between the two quarks formed in the
subprocess $\gamma^* g^*\rightarrow q\bar{q}$ \cite{22,23,24,25} and follows from the angular 
ordering condition. Processes different in their topology may allow different forms
of $\mu^2_F$.

On the other hand, one can argue that the definition of $\mu^2_F$ must preserve
strict consistency between the fitting procedure (from which the gluon densities 
are obtained) and the actual calculation (to which the gluon densities are applied) 
and, therefore, have the form $\mu_F\propto\hat{s}+\mathbf{Q}^2_T$ (probably with some
numerical prefactor). 
The property that the parameter space ($\mu^2_F$ versus $\mathbf{k}^2_T$) 
is one-dimensional rather than two-dimensional is not dangerous by its own 
(and is similar to collinear factorization). The choice of $\mu_F$ is not free, 
but is part of the proposed TMD parametrization. We simply have to take the gluon 
densities obtained from the fit (with the given form of $\mu_F$) as they are and 
directly substitute them into actual calculations.

An interesting parallel can be seen with collinear calculations at the NLO \cite{48}. 
As one can see from Figs.~6 and~7 in that paper, the integral size 
of the NLO contribution exceeds the LO result by a significant (huge) factor. 
The difference between the NLO 
and LO predictions is most dramatic in the region of small $J/\psi$ invariant masses and,
respectively, small rapidity separation. In the $k_T$-factorizaion approach, the NLO
contributions are absorbed into the evolution of TMD gluon densities, where the parameter
$\mu_F$ regulates the gluon emission. So, we find it not surprising that the effect
of $\mu_F$ is largest in the same region where the NLO contribution is largest, i.e.
at small $m(J/\psi, J/\psi)$ and small $\Delta y(J/\psi, J/\psi)$. This has to be compared
with our Fig. 7. Were the authors of \cite{48} had calculated the $p_T(J/\psi, J/\psi)$ 
distributions, they would undoubtedly observe the same discrepancy between NLO and LO 
as we do in Fig.~7 bottom right. We had, in due time, reproduced all of the results \cite{48} with our $k_T$-factorizaion technique.

We would not, however, deny the fact that the numerical instability at
$\mu^2_F \simeq \mathbf{k}^2_T$ indicates that we are approaching the applicability limits 
of the $k_T$-factorization (non-collinear parton evolution).
The problem of extraordinary sensitivity of the results to the choice of $\mu_F$
has probably been first detected in \cite{49}, but left without further attention and analysis.


\section{Numerical results} \indent

In this section, we present the results of our calculations 
and perform a comparison with the recent ATLAS data 
collected at $\sqrt{s}=8$ TeV \cite{20}. The ATLAS Collaboration has measured the differential cross sections of 
prompt $J/\psi$ pair production as functions of the transverse momentum $p_T(J/\psi,J/\psi)$ and the invariant
mass $m(J/\psi,J/\psi)$ of the $J/\psi$ pair, of the rapidity separation $\Delta y(J/\psi,J/\psi)$ 
and of the azimuthal angle difference $\Delta \phi (J/\psi,J/\psi)$ between the two $J/\psi$ mesons. 
The following selection criteria were applied: $p_T(J/\psi) > 8.5$~GeV, $|y(J/\psi)| < 2.1$, 
$|\eta(\mu)| < 2.3$ for all muons, $p_T(\mu) > 4$~GeV for the muons from the triggered $J/\psi$ meson,
and $p_T(\mu) > 2.5$~GeV for the muons from the other $J/\psi$ meson.  Also presented were the 
differential cross sections as functions of the subleading 
$J/\psi$ transverse momentum $p_T(J/\psi_2)$, of the $J/\psi$ pair transverse momentum, and of the
$J/\psi$ pair invariant mass, with the selection criteria $p_T(J/\psi) > 8.5$~GeV, $|y(J/\psi)| < 2.1$ 
for each $J/\psi$, without requirements on the muons in the final state. 
These measurements were performed in the central $|y(J/\psi_2)| < 1.05$ and forward $1.05 < |y(J/\psi_2)| < 2.1$ 
rapidity intervals. We have implemented the same selection criteria in our calculations. 

The calculations are performed with the following setting of hard scales. 
For the CS contributions (\ref{CS-1})--(\ref{CS-2}) we take 
$\mu^2_F = \frac{1}{4}(\hat{s} + \mathbf{Q}^2_T)$ 
and $\mu^2_R = \frac{1}{4}(m^2_{\psi} + \frac{1}{2}(p^2_{1T} + p^2_{2T}))$
with $p_{1T}$ and $p_{2T}$ being the transverse momenta of produced particles. 
For these processes we use the gluon densities JH'2013 set 1 muf$-$ and JH'2013 set 2 muf$-$
obtained with $\mu^2_F = \frac{1}{4}(\hat{s} + \mathbf{Q}^2_T)$.
So, the numerical prefactor 1/4 in the definition of $\mu_F$ holds exact correspondence
with the fitting procedure \cite{45}; and, in comparison with the choice
$\mu^2_F = (\hat{s} + \mathbf{Q}^2_T)$, it softens to some extent the scale dependence
of the results.

For the fragmentation contributions and DPS we set 
$\mu^2_F = \hat{s} + \mathbf{Q}^2_T$, $\mu^2_R = m^2_{\psi} + p^2_T$ 
for subprocesses with outgoing off-shell gluon or bound charmed pair;
and $\mu^2_F = \hat{s} + \mathbf{Q}^2_T$, 
$\mu^2_R = m^2_{c} + \frac{1}{2}(p^2_{1T} + p^2_{2T})$ 
for subprocesses with outgoing unbound charmed quarks. 
The factorization concept implies that the fragmentation scale can only depend on the 
parameters of the fragmented parton and must not depend on the hard interaction scale. 
So, we set $\mu^2_{\rm frag} = m^2_Q + p^2_T$ where $m_Q$ and $p_T$ are the mass and 
transverse momentum of the fragmented parton.

\begin{figure}
\begin{center}
{\includegraphics[width=.48\textwidth]{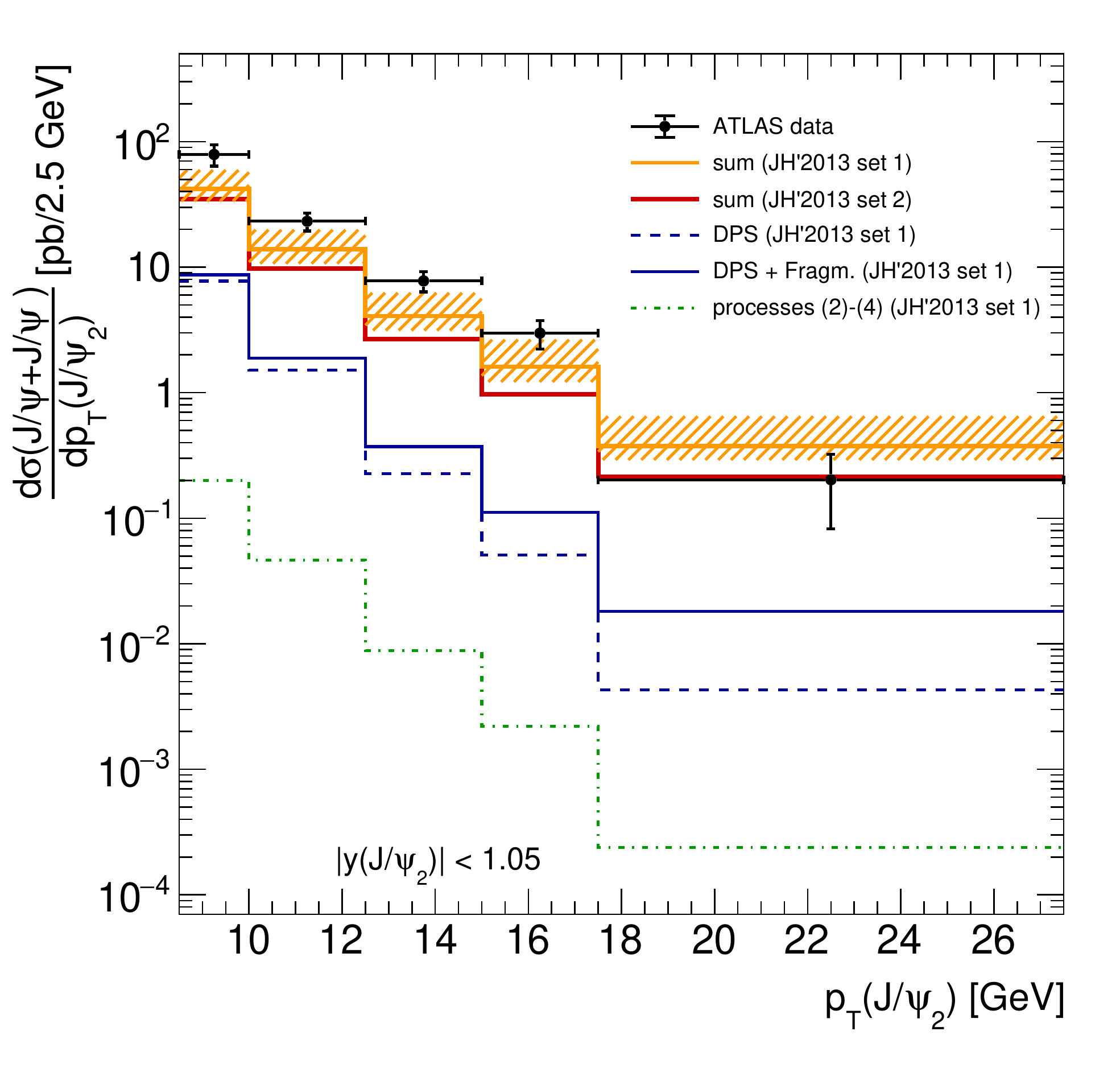}}\hfill
{\includegraphics[width=.48\textwidth]{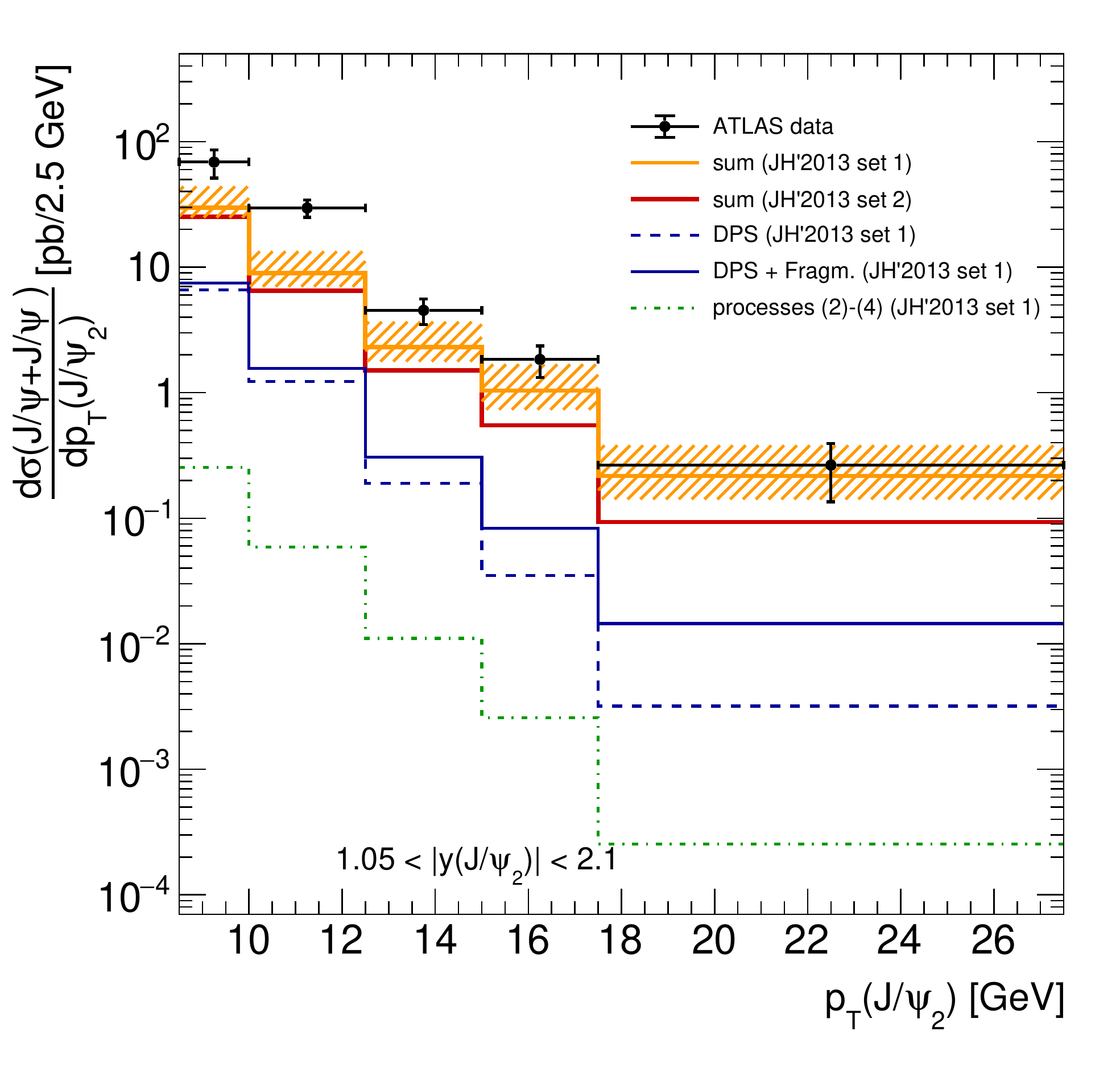}}\hfill
{\includegraphics[width=.48\textwidth]{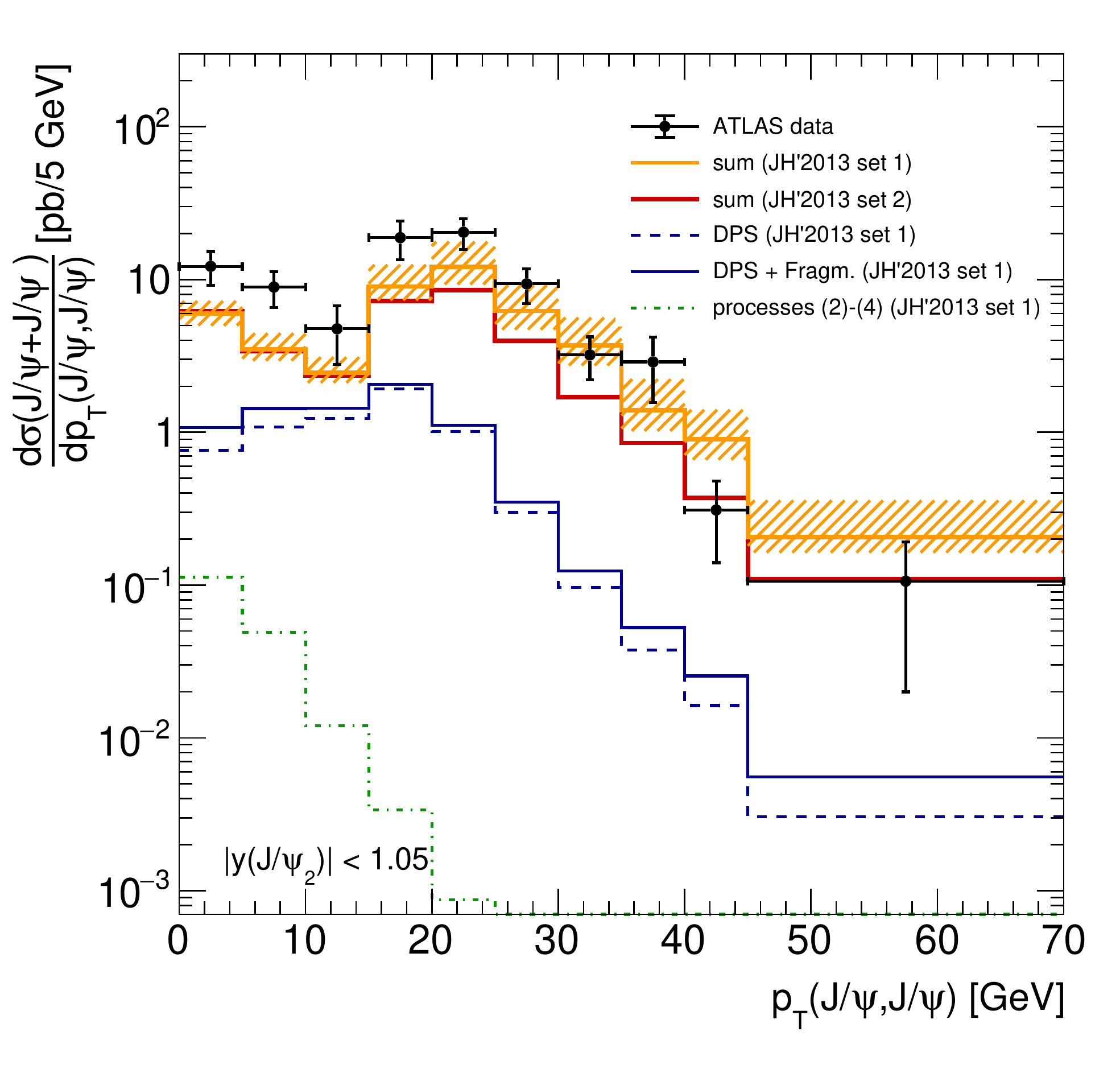}}\hfill
{\includegraphics[width=.48\textwidth]{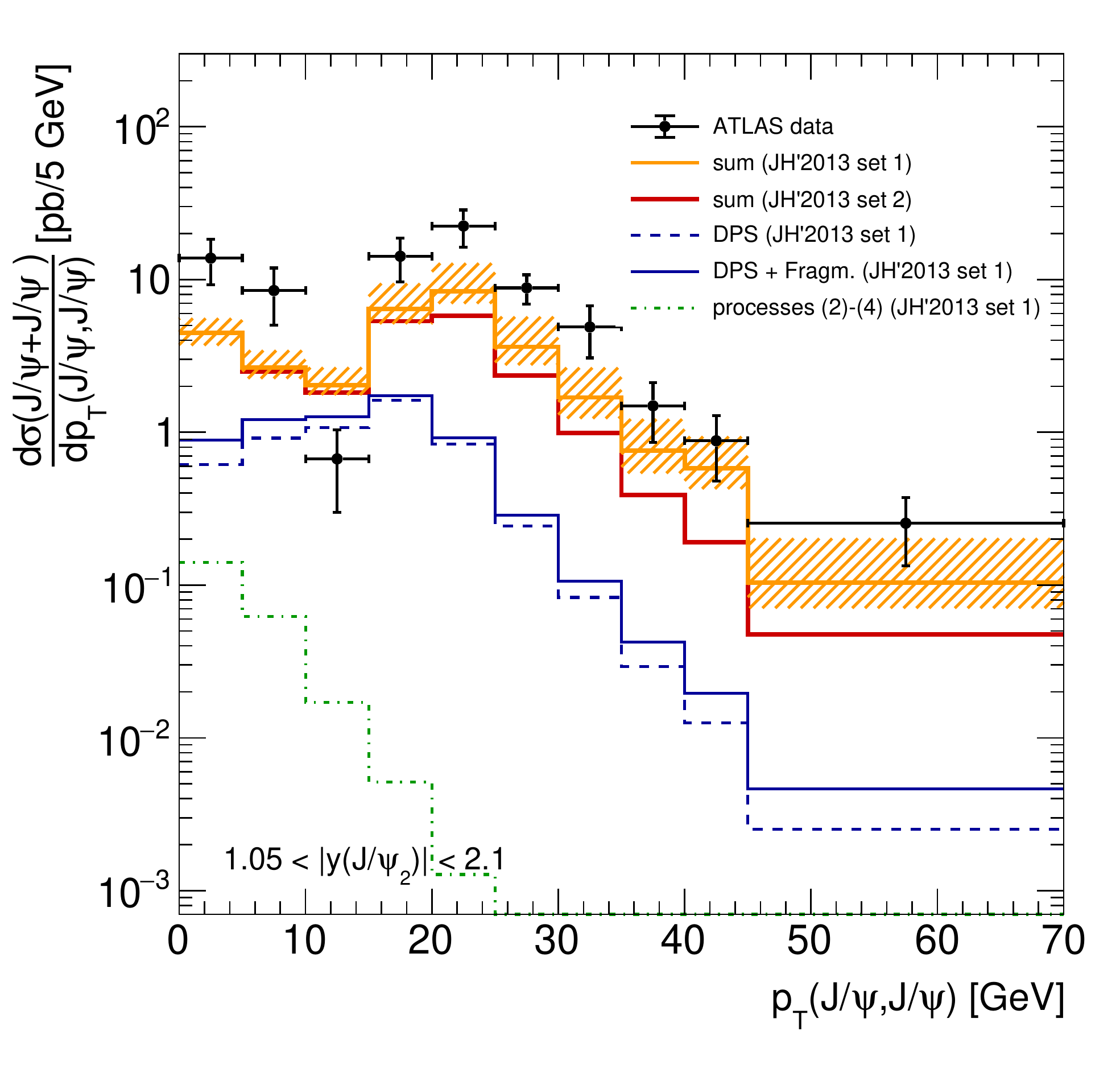}}\hfill
{\includegraphics[width=.48\textwidth]{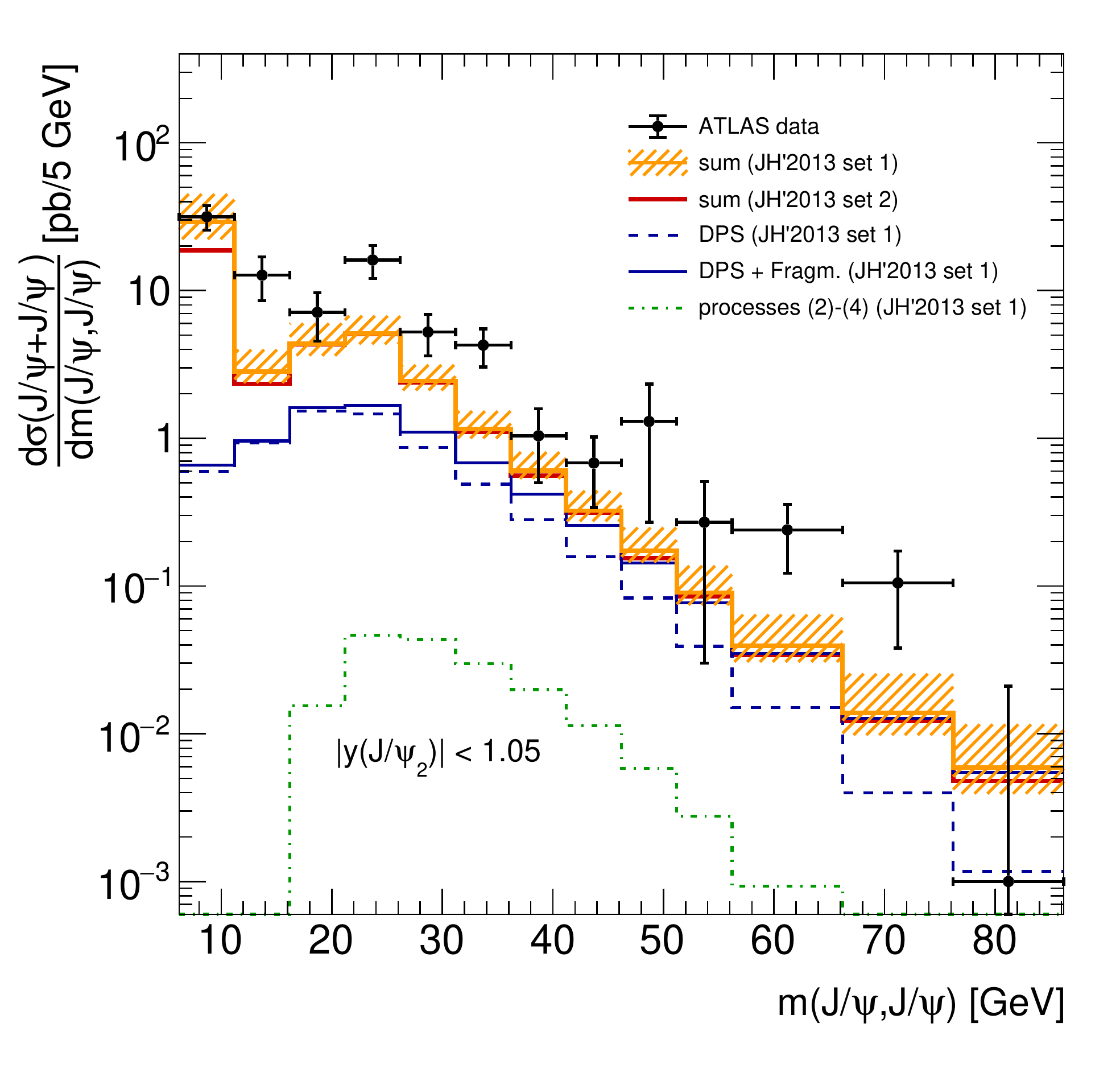}}\hfill
{\includegraphics[width=.48\textwidth]{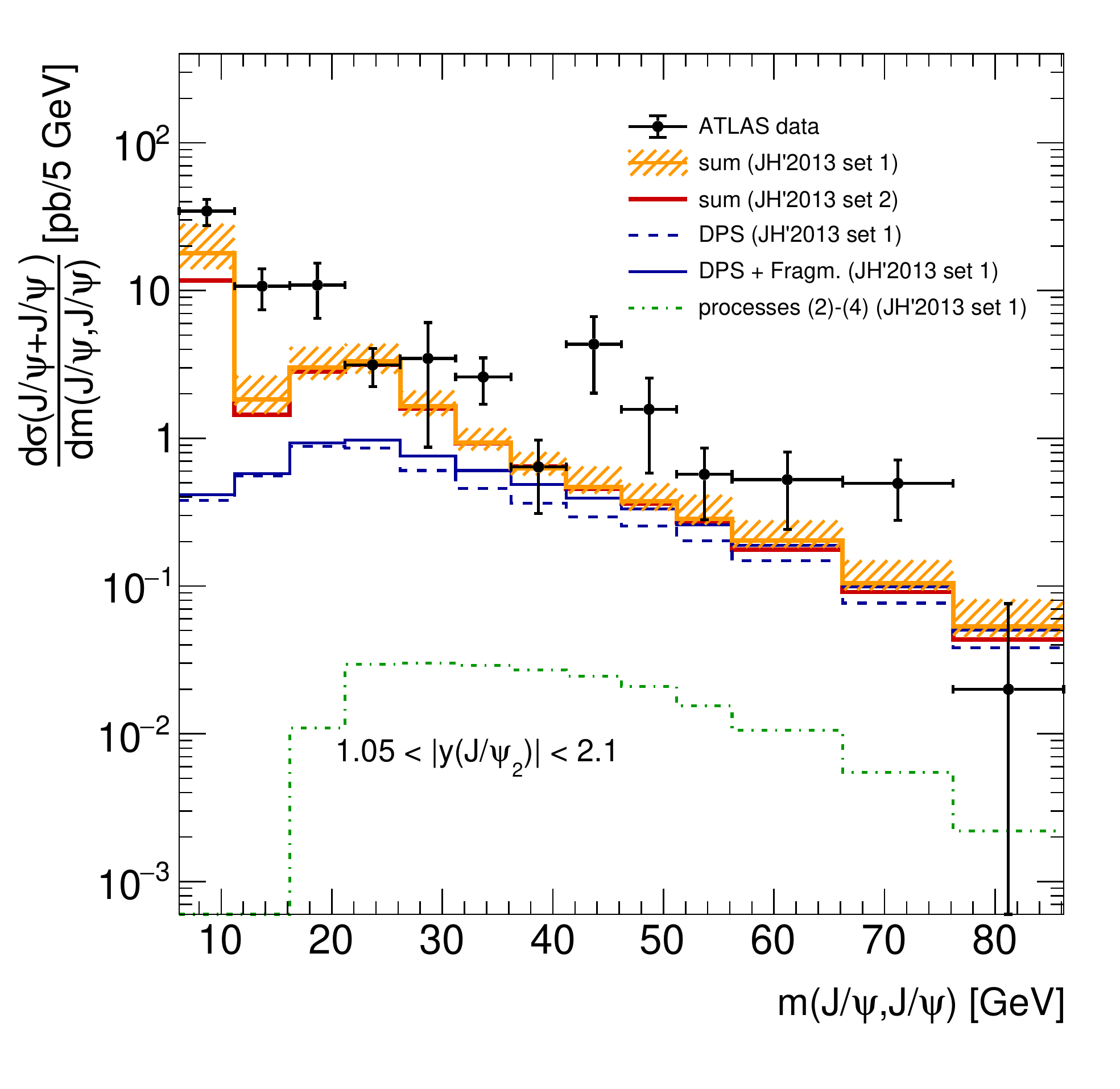}}\hfill

\caption{
The differential cross-section of double $J/\psi$ production in $pp$ collisions at $\sqrt{s} =$ 8 TeV 
in the central (left column) and forward (right column) rapidity regions, as a function of subleading $J/\psi$ transverse momentum, $p_T(J/\psi_2)$ (upper row); $J/\psi$ pair transverse momentum,
 $p_T(J/\psi,J/\psi)$ (middle row);  $J/\psi$ pair invariant mass, $m(J/\psi,J/\psi)$ (lower row). 
The integral theoretical predictions are shown for JH'2013 set 1 and JH'2013 set 2 as well as the 
individual DPS and fragmentation contributions. The experimental data are taken from \cite{20}.
}
\label{fig:ATLAS_plot1}
\end{center}
\end{figure}

\begin{figure}
\begin{center}
{\includegraphics[width=.5\textwidth]{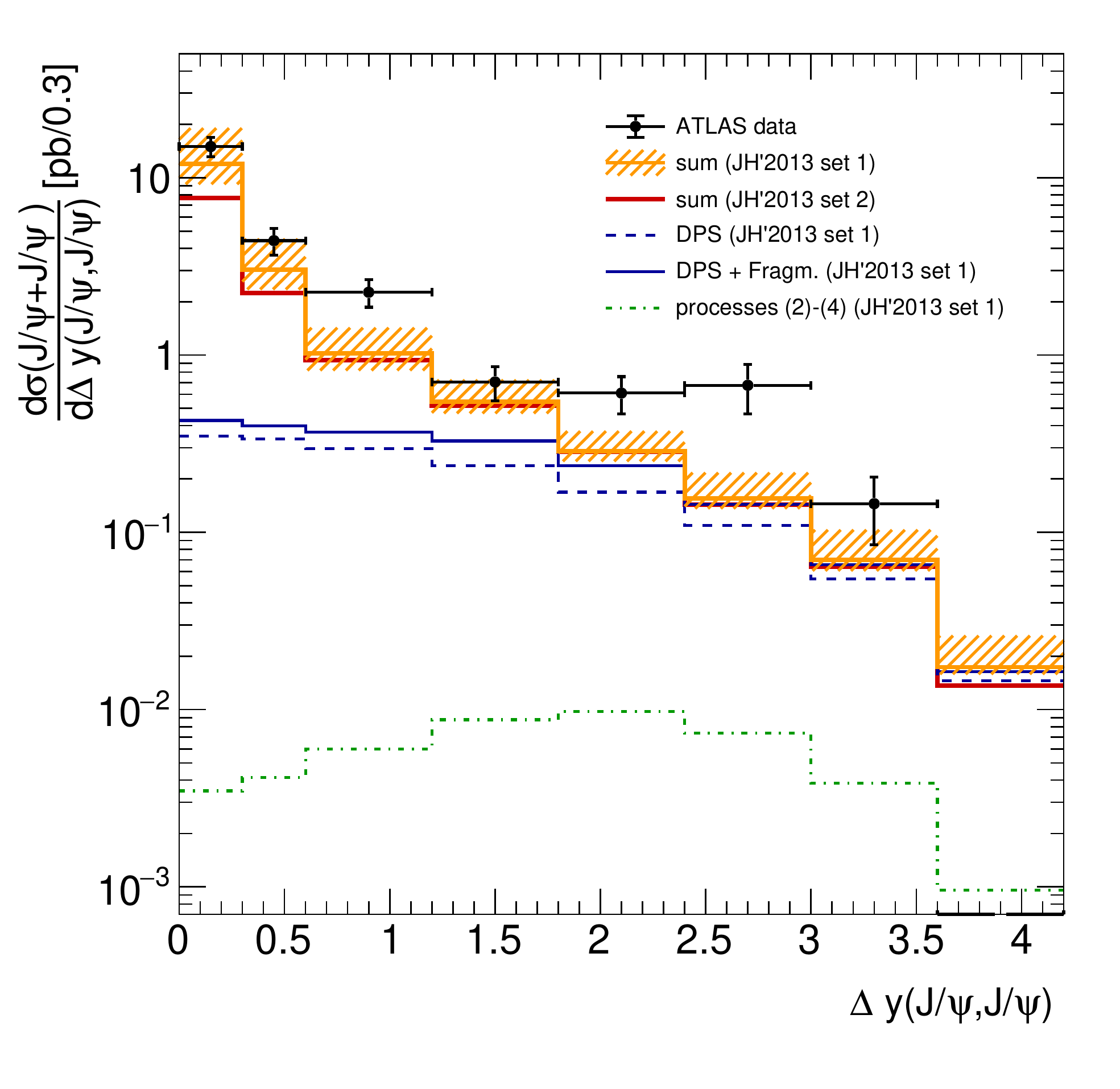}}\hfill
{\includegraphics[width=.5\textwidth]{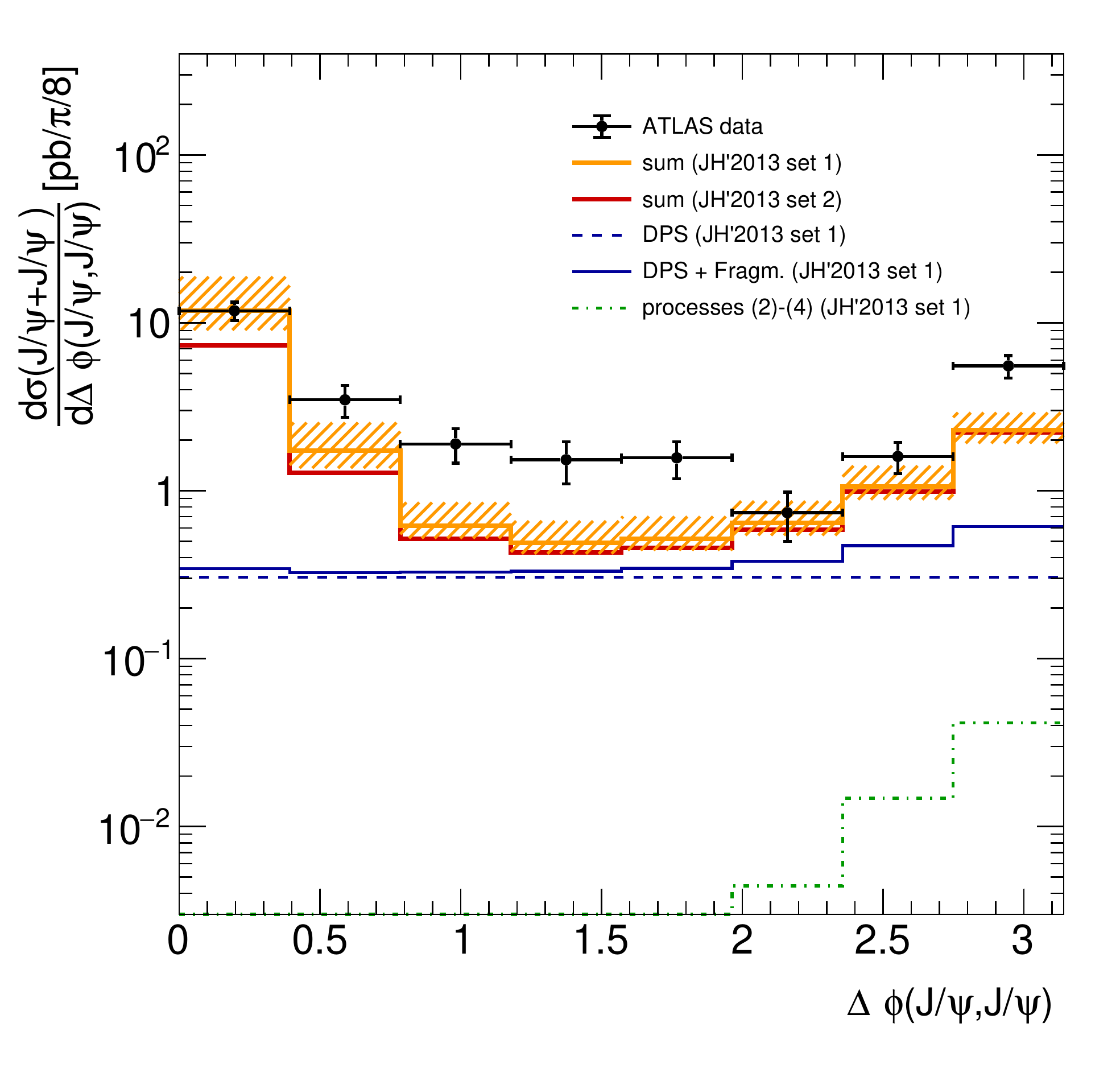}}\hfill
\\
{\includegraphics[width=.5\textwidth]{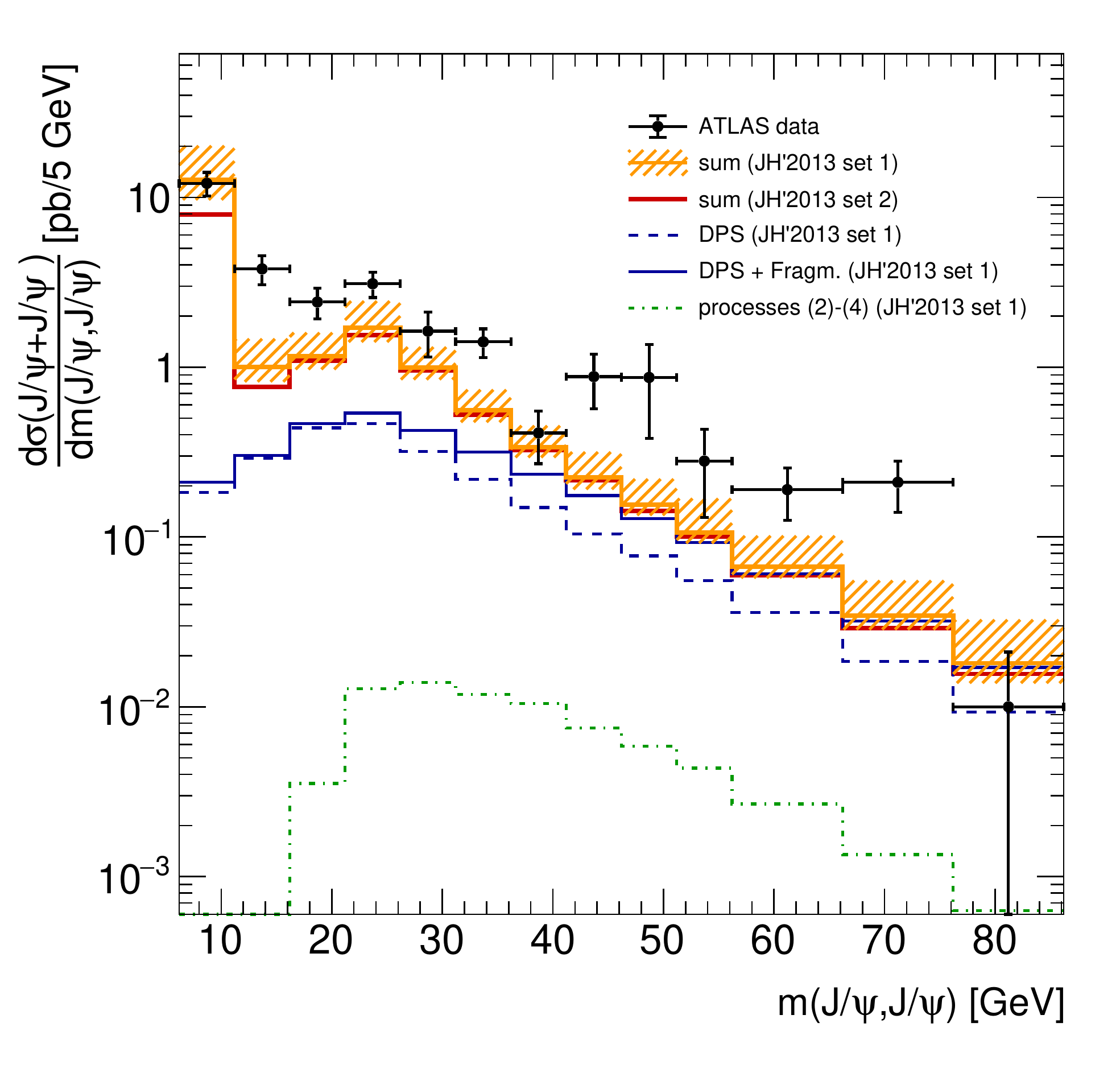}}\hfill
{\includegraphics[width=.5\textwidth]{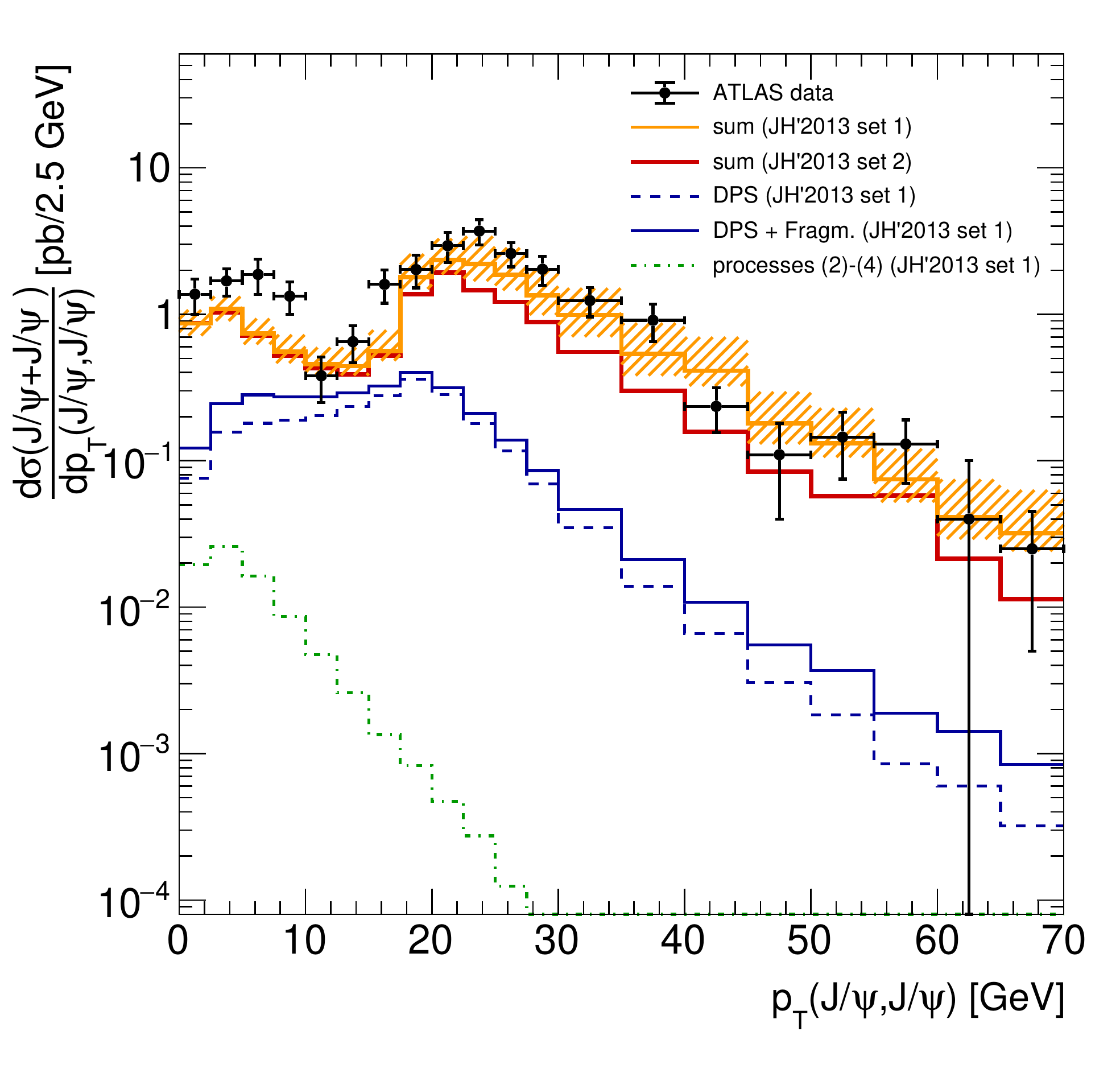}}\hfill

\caption{
The differential cross-section of double $J/\psi$ production in $pp$ collisions at $\sqrt{s} =$ 8 TeV 
 as a function of rapidity separation, $\Delta y(J/\psi,J/\psi)$ (left upper panel); 
 azimuthal angle difference, $\Delta \phi(J/\psi,J/\psi)$ (right upper panel); 
 $J/\psi$ pair invariant mass, $m(J/\psi,J/\psi)$ (left lower  panel); 
 $J/\psi$ pair transverse momentum, $p_T(J/\psi,J/\psi)$ (right lower panel). 
The integral theoretical predictions are shown for JH'2013 set 1 and JH'2013 set 2 as well as the
individual DPS and DPS + fragmentation contributions. The experimental data are taken from \cite{20}.
}
\label{fig:ATLAS_plot2}
 \end{center}
\end{figure}

Our numerical results are shown in Figs.~\ref{fig:ATLAS_plot1} and~\ref{fig:ATLAS_plot2}.
One can see that the predictions obtained with the JH'2013 set 1 gluon density 
(used as the default choice) are rather close to the data and can be said compatible 
with them within theoretical uncertainties (shaded orange bands) for the majority of the 
measured distributions. 
As usual, the theoretical uncertainties were estimated by varying the renormalization 
scales around their default values by a factor of $2$. 

To highlight the role of DPS mechanism and the combinatorial effects of multiple gluon
and/or quark radiation in the initial state, 
we separately show the relevant contributions.
The blue dashed histograms in Figs.~\ref{fig:ATLAS_plot1} and~\ref{fig:ATLAS_plot2}
correspond to the estimated DPS terms, whereas
the blue solid histograms (labeled as "DPS + Fragm.")
represent the sum of DPS and fragmentation contributions.
As one can see, the effect of multiple parton radiation is very essential in the region
of high invariant mass $m(J/\psi,J/\psi)$ and large rapidity separation $\Delta y(J/\psi,J/\psi)$. 
An accurate account of these contributions is extremely important for ATLAS data.
The importance of fragmentation terms have been already demonstrated earlier \cite{21},
where, however, only the $g^* \to cc[^3S_1^{(8)}]$ transition has been taken into account.
Our present calculations include a much larger  number of fragmentation channels; 
and the feeddown contributions from the $\chi_c$ and $\psi^\prime$ decays are
also taken into account.
Finally, we show the contributions from subprocesses~(\ref{CS-2-1}) --- (\ref{CS-2})
(blue solid curves), although 
we find these subprocesses not playing a considerable role for any observable.

To investigate the sensitivity of our results to the choice of TMD gluon density, we 
have repeated our calculations with JH'2013 set 2 distribution. We find that the latter 
predictions (red histograms in Figs.~\ref{fig:ATLAS_plot1} and~\ref{fig:ATLAS_plot2})
are in good agreement with the ones obtained with the default JH'2013 set 1 gluon density
for the majority of observables. There are some discrepancies in the region of high 
$p_T(J/\psi,J/\psi)$ and $p_T(J/\psi_2)$. They can be promptly attributed to the different
behavior of TMD gluon distributions at high transverse momenta $k_T$. 

Next, we discuss the role of multiple gluon radiation in the DPS contributions.
As it was said above, our calculation scheme implies that the direct CO subprocesses
$g^* + g^* \to c\bar{c}[^3S^{(8)}_1]$ in the both hard interaction blocks have to be 
replaced with the $g^* + g^* \to g^*$ and/or $g^* + g^* \to c + \bar{c}$ subprocesses 
accompanied by gluon evolution ladders (reconstructed according to the CCFM equation)
with subsequent fragmentation of all emitted partons into charmonia.
This brings additional contributions to the DPS production cross section.
To show the role of these terms in more detail, we separately consider three different
sources\footnote{Here we will not consider feeddown contributions to double $J/\psi$ production.}. 
The first of them 
contains only $[^3S^{(1)}_1,\,^3P^{(8)}_J] \times [^3S^{(1)}_1,\,^3P^{(8)}_J]$ combinations
of single $J/\psi$ production mechanisms.
The second source includes $[^3S^{(8)}_1] \times [^3S^{(1,8)}_1,\,^3P^{(8)}_J]$ subprocesses.
The sum of these two sources gives the DPS contribution to 
double $J/\psi$ production in the conventional scheme.
The third source is the modification of the second one: it includes the
$[g,\,c] \times [g,\,c,\,^3S^{(1)}_1,\,^3P^{(8)}_J]$
combinations, where the symbols $g$ and $c$ denote fragmentation contributions coming
from the $g^* + g^* \to g^*$ and $g^* + g^* \to c + \bar{c}$
subprocesses (with multiple gluon radiation taken into account in both cases). The sum 
of the first and the third contributions represents the DPS cross section in our scheme.

\begin{figure}
\begin{center}
{\includegraphics[width=.5\textwidth]{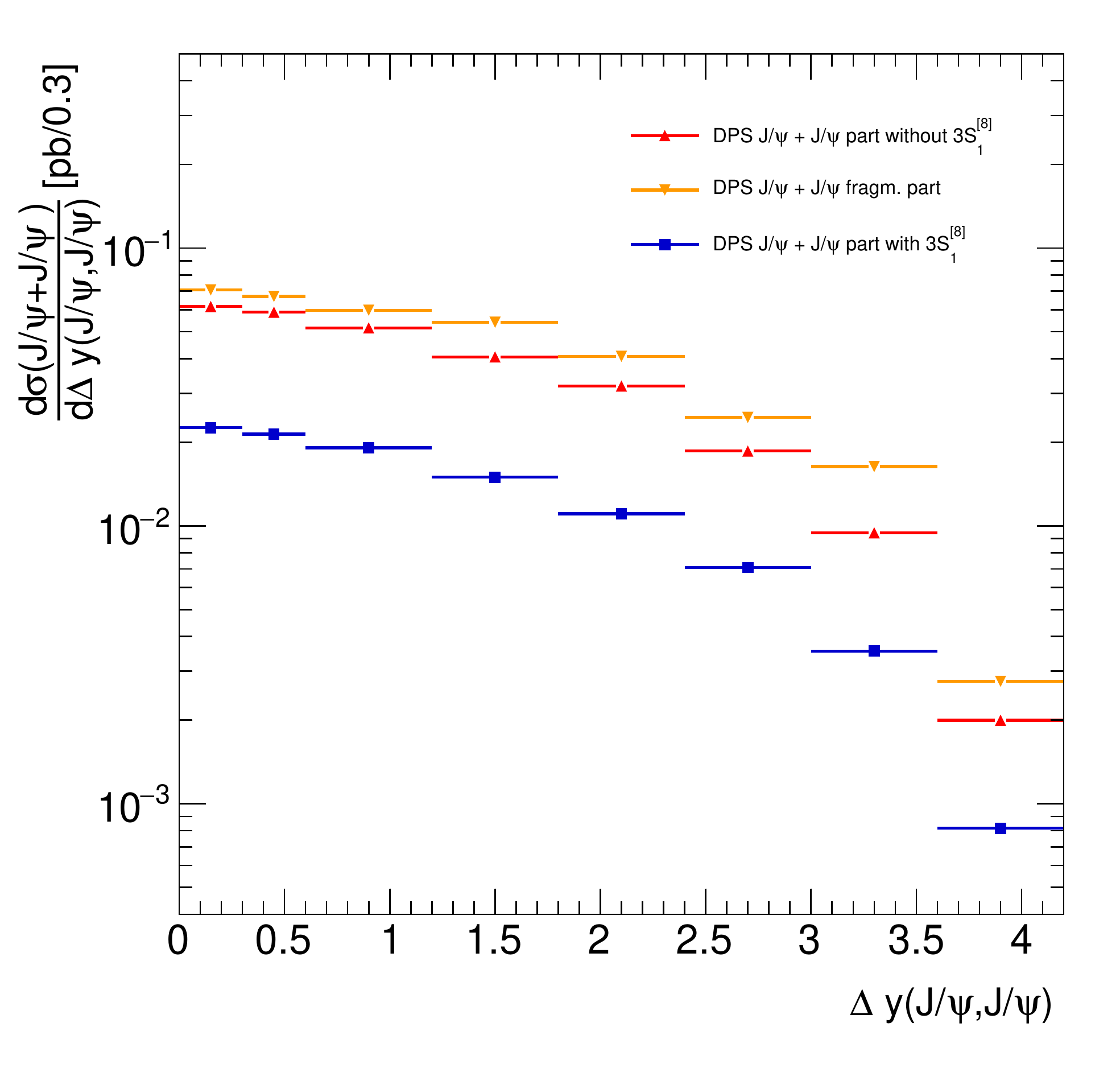}}\hfill
{\includegraphics[width=.5\textwidth]{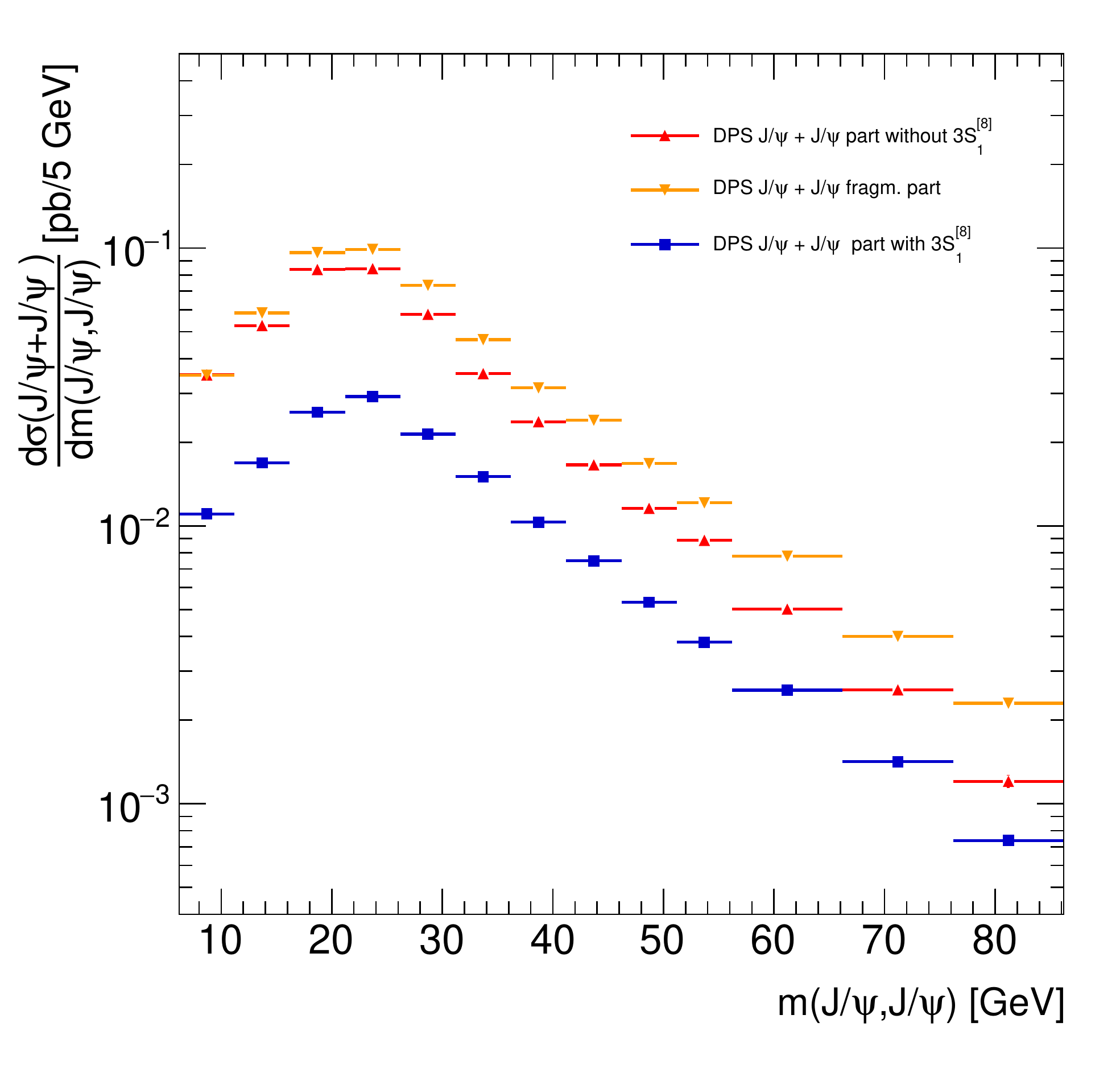}}\hfill
\caption{DPS contributions to the direct $J/\psi$ pair production cross-section
as a function of rapidity separation between the two mesons, $\Delta y(J/\psi,J/\psi)$ 
(left panel) and of the $J/\psi$ pair invariant mass, $m(J/\psi,J/\psi)$ (right panel).
The modified and the conventional contributions are shown separately. 
}
\label{fig:DPS_details}
 \end{center}
\end{figure}

A comparison between the different contributions obtained within the DPS framework is 
displayed in Fig.~\ref{fig:DPS_details}. It is clearly seen that the yield from modified 
mechanism (i.e., the third source, orange lines) is approximately three times greater 
than that from the conventional mechanism (the second source, blue lines). Eventually, 
it nearly doubles the total estimated DPS contribution. This constitutes in our view 
a remarkable result.
Thus, an accurate treatment of multiple gluon emission in the initial state is indeed
very important for evaluating the DPS cross sections. 
In fact, these additional contributions could also play a role for a number of other
processes related to the multiparton interactions. This can, in turn, motivate some
revisions of the effective DPS cross section $\sigma_{\rm eff}$ extracted from the 
experimental data.

\section{Conclusion} \indent 

In the present study, we have addressed a challenging problem of prompt double $J/\psi$ 
production in $pp$ collisions at the LHC conditions.
In addition to the conventional production mechanisms mentioned earlier in the literature,
we take into account the effects of multiple gluon radiation in the initial state 
followed by gluon fragmentation into $J/\psi$ mesons. The evolution of the radiated
gluon cascade is described within the $k_{T}$-factorization approach, with making use
of CCFM equation. We demonstrate the importance of these new contributions both in single 
and double parton scattering processes.

We paid much attention to inspecting the influence of the factorization scale 
$\mu_F$ on the numerical errors and to studying the kinematic conditions under which 
the computations become unstable. We find that the dangerous region is $p_T^2 \gg\hat{s}$,
that corresponds to $\mu_F^2\simeq {\mathbf k}_T^2$.
The computations can be made stable by adopting either of two alternatives.
One can adopt just a different definition of the scale $\mu_F$, so that to make it 
far from the instability region (basically, by taking $\mu_F$ independent on $k_T$).
As an alternative, one can consider the definition $\mu_F\propto\hat{s}+\mathbf{Q}^2_T$
as a built-in property of the TMD gluon density.
The latter choice preserves the correspondence between the TMD fitting procedure and 
the actual calculations where the obtained TMD gluon distribution is used.

In the framework of our second hypothesis, we have performed a numerical comparison with the latest measurements reported by the ATLAS collaboration
for  $\sqrt{s} = 8$~TeV.
Having the fragmentation mechanism employed in the theory, we greatly reduce
the discrepancy between the predictions and the data.
This has immediate impact on the value of the effective cross section 
$\sigma_{\rm eff}$ that parametrises the DPS contribution.

\section*{Acknowledgements} \indent

We thank M.A.~Malyshev, G.I.~Lykasov and H.~Jung for their important comments and remarks.
This research was supported by the Russian Science Foundation under grant 22-22-00119.
A.A.P. was also supported in part by the RFBR grant No. 20-32-90105.

\end{document}